\documentclass[a4paper,11pt]{article}

\usepackage[a4paper,left=2.73cm,right=2.7cm,top=3cm,bottom=3.5cm]{geometry}
\usepackage{amsmath,amssymb,graphicx,colortbl,xcolor,caption,subcaption,wrapfig}
\usepackage[colorlinks=true,linktocpage=true,linkcolor=blue,citecolor=blue]{hyperref}

\usepackage[all]{xy}
\definecolor{gray}{rgb}{.9,.9,.9}

\usepackage{caption}
\usepackage{subcaption}

\newcommand{\dd}{\mathrm{d}}

\newcommand{\mt}[1]{\textrm{\tiny #1}}
\newcommand{\nc}{N_\textrm{c}}
\newcommand{\ncc}{n_\textrm{c}}
\newcommand{\Qf}{Q_\textrm{f}}
\newcommand{\Qc}{Q_\textrm{c}}
\newcommand{\qc}{q_\textrm{c}}

\newcommand{\QQc}{\mathcal{Q}_\textrm{c}}
\newcommand{\uf}{U_{\textrm{\tiny flavor}}}
\newcommand{\mq}{m_{\rm q}}

\newcommand{\fig}[1]{Fig.~\ref{#1}}

\newcommand{\nf}{N_\textrm{f}}
\newcommand{\nff}{n_\textrm{f}}

\newcommand{\eqq}[1]{(\ref{#1})}

\newcommand{\be}{\begin{equation}}
\newcommand{\ee}{\end{equation}}
\newcommand{\sac}{\, , \qquad}

\newcommand{\bea}{\begin{eqnarray}}
\newcommand{\eea}{\end{eqnarray}}

\definecolor{gray}{rgb}{.9,.9,.9}

\makeatletter \@addtoreset{equation}{section}
\makeatother

\begin{document}

\begin{titlepage}

\hfill{ICCUB-16-028}

\vspace{1cm}
\begin{center}

	    { \LARGE{\bf Unquenched flavor on the Higgs branch}}
		
\vskip 45pt

{\large \bf Ant\'on F. Faedo,$^{1}$ David Mateos,$^{1,2}$ Christiana Pantelidou,$^1$ 
\\ and 
Javier Tarr\'\i o$^{3}$}
		
		\vspace{25pt}

		{$^{1}$ Departament de F\'\i sica Qu\`antica i Astrof\'\i sica \&  Institut de Ci\`encies del Cosmos (ICC), Universitat de Barcelona, Mart\'{\i}  i Franqu\`es 1, 08028 Barcelona, Spain.}\\
		\vspace{15pt}
		{$^{2}$Instituci\'o Catalana de Recerca i Estudis Avan\c cats (ICREA), \\
		Passeig Llu\'\i s Companys 23, 08010 Barcelona, Spain.}\\
		\vspace{15pt}
		{$^{3}$ Physique Th\'eorique et Math\'ematique, Universit\'e Libre de Bruxelles\\
		and International Solvay Institutes, ULB-Campus Plaine CP231, B-1050 Brussels, Belgium.}
		
\vskip 10pt
\end{center}

\vspace{10pt}
\abstract{\normalsize
We construct  the gravity duals of the Higgs branches of three-dimensional (four-dimensional) super Yang--Mills theories coupled to $\nf$ quark flavors. The effect of the quarks on the color degrees of freedom is included, and corresponds on the gravity side to the  backreaction of $\nf$ flavor D6-branes (D7-branes) on the background of $\nc$ color D2-branes (D3-branes). The Higgsing of the gauge group arises from the dissolution of some color branes inside the flavor branes. The dissolved color branes are represented by non-Abelian instantons whose backreaction is also included. The result is a cascading-like solution in which the effective number of color branes varies along the holographic direction. In the three-dimensional case the solution may include an arbitrary number of quasi-conformal (walking) regions. 

}

\end{titlepage}

\tableofcontents

\hrulefill
\vspace{15pt}

\section{Introduction} 

The low-energy worldvolume dynamics of a collection of $\nc$ Dp-branes in flat space is 
described by maximally supersymmetric Yang--Mills theory (SYM) in \mbox{$d=p+1$} dimensions with gauge group $SU(\nc)$. The moduli space of this theory (with $2\leq p\leq 7$) is a Coulomb branch parametrized by the expectation values of the scalar fields in the adjoint representation of the gauge group. In the D-brane realization the origin of this  moduli space corresponds to the situation in which all the branes are coincident, and it can be described holographically by a gravitational solution with a single throat. In contrast, a generic point on the moduli space corresponds to non-coincident branes and is holographically modelled by a geometry with multiple throats \cite{Klebanov:1999tb}. 

If the above SYM theory is coupled to matter in the fundamental representation in  a supersymmetry-preserving way, then the resulting theory possesses a Higgs branch parametrized by mesonic-like operators. We will refer to the fundamental matter as `flavor' and, although it includes both bosonic and fermionic degrees of freedom, also  as `quarks'. In the D-brane realization, the flavor degrees of freedom can be added by considering an appropriate intersection of the original $\nc$ `color' branes and a collection of $\nf$ `flavor' D(p+4)-branes. The two cases of interest for us will be the D2-D6 and the D3-D7 systems, respectively dual to three- and four-dimensional SYM theory with quarks. The Higgs branch is then described by a situation in which some of the color branes have dissolved inside the flavor branes. From the viewpoint of the flavor branes, the dissolved color branes are described by an instantonic configuration of the non-Abelian  gauge field living on their worldvolume \cite{Witten:1995gx,Douglas:1995bn}.

The holographic description of the Higgs branch has been studied in the so-called 
``probe approximation''. Early work includes \cite{Erdmenger:2005bj,Guralnik:2004ve}; Ref.~\cite{Arean} will be particularly useful to us. In this approximation the backreaction on the spacetime metric and the other supergravity fields of the flavor branes and the instanton is neglected. This is justified if the large-$\nc$ limit is taken \`{a} la 't Hooft, i.e.~with $\nf$ fixed so that $\nf/\nc \to 0$. In the gauge theory this corresponds to a ``quenched'' approximation in which the effect of the quarks on the dynamics of the gluons and the adjoint matter is ignored. In this approximation the most dramatic property of the Higgs branch, namely the change of the effective rank of the gauge group as a function of the energy scale, is not visible. The purpose of this paper is to construct the holographic description of the Higgs branch in the unquenched case, namely to include the backreaction of the flavor branes and the instanton. In particular, this means that we are taking the large-$\nc$ limit \`{a} la Veneziano, i.e.~keeping $\nf/\nc$ non-zero.  As we will see, in our solutions the Higgsing of the gauge group results in a cascading-like solution in which the effective number of color branes varies as a function of the holographic radial coordinate. 

The starting point for our construction will be the $\mathcal{N}=1$ supersymmetric solutions found in \cite{Faedo} and \cite{Benini}. These provide the holographic description of three- and four-dimensional SYM  theories with flavor, and of some quiver-like generalizations thereof, at the origin of their moduli spaces. 
In the setups of \cite{Faedo,Benini} each of the $\nf$ individual flavor branes wraps a different cycle of the internal geometry. The branes are smeared or distributed around the internal manifold by the action of a (subset of) its symmetry group. The number of branes $\nf$ is taken to be large so that the distribution can be considered continuous on scales relevant to the supergravity description, but in such a way that the distance between branes is still large compared to the string scale. In this way the action for the entire set of flavor branes consists simply of $\nf$ copies of the Abelian action for a single brane. In other words, the flavor group is effectively $\nf$ copies of U(1). The smearing procedure (see \cite{Nunez:2010sf} for a review) vastly simplifies the construction at the technical level, since it reduces the equations of motion to a set of ordinary differential equations. 

In order for the gauge theory to possess a Higgs branch, however, we must have a non-Abelian flavor group U($\nff$).\footnote{See however \cite{Ammon} for an ``Abelian Higgs branch'' in the presence of a chemical potential.} 
To describe this on the gravity side we will consider $\nf/\nff$ stacks of flavor branes with $\nff$ coincident  branes each, instead of $\nf$ individual branes. In this way the effective action for the flavor branes becomes $\nf/\nff$ copies of the non-Abelian action for $\nff$ overlapping D-branes. Each stack of flavor branes will carry a U($\nff$) instanton solution on its worldvolume, and the branes+instanton system will then be smeared over the internal geometry. 

In Sec.~\ref{sec.3dgaugetheories} we focus on three-dimensional gauge theories. This part of the paper discusses the holographic description of the Higgs branch of supersymmetric flavored Yang--Mills, Chern--Simons-Matter and quiver type theories. In this case most of the solutions  are completely analytic and can be written in closed form.  
We discuss both massless and massive quarks and comment  on the implementation of our ideas for ABJM-type theories. 

Sec.~\ref{sec.4dgaugetheories} discusses flavored Yang--Mills and  Klebanov--Witten quiver gauge theories in four  dimensions. In this case we focus on massless quarks only. This section builds on the intuition developed in Sec.~\ref{sec.3dgaugetheories} and requires the use of some numerics in order to construct the solutions. 

In Sec.~\ref{sec.discussion} we summarize our results.

\section{Three-dimensional gauge theories}\label{sec.3dgaugetheories}

The starting point for constructing the gravity dual of the Higgs branches of 
$\mathcal{N}=1$ supersymmetric three-dimensional gauge theories  with fundamental matter are the solutions found in \cite{Faedo}.  These provide the holographic description of these theories at  the origin of their moduli spaces. The six-dimensional internal geometries of the solutions are nearly K\"ahler manifolds (NK). If the NK manifold  is taken to be a six-sphere then the dual gauge theory is SYM theory. Otherwise the gauge theory is a quiver-like theory. 

The NK property implies the existence of a real two-form $J$, and a complex three-form $\Omega$, both globally-defined and verifying the differential conditions
\be
\dd J\,=\,3\,{\rm Im}\,\Omega\,,\qquad\qquad\qquad\dd{\rm Re}\,\Omega\,=\,2\,J\wedge J\,,
\label{dif}
\ee
and the wedgings 
\be
J\wedge\Omega\,=\,0\,,\qquad\qquad\qquad\frac13J\wedge J\wedge J\,=\,\frac{i}{4}\Omega\wedge\overline{\Omega} \,.
\ee
For later convenience we choose the orientation to be such that 
\be
J\wedge J\wedge J=-6\,\omega_6 \,, 
\label{vol}
\ee
with $\omega_6$ the volume form. The forms $J$ and  $\Omega$ will play an important role in our construction.

On the gravity side, the flavor degrees of freedom are represented by sets of D6-branes in the geometry sourced by the color D2-branes. The flavor branes come in  stacks of $\nff$ branes each, so we will start with the non-Abelian action for $n_{\rm f}$ coincident D6-branes. Although we adopt  the symmetrized-trace prescription \cite{Myers:1999ps}, the result would be the same with other prescriptions since we will only consider supersymmetric configurations. The Dirac--Born--Infeld (DBI) part of the action in string frame takes the form
\begin{equation}\label{nADBI}
S_\textrm{\tiny DBI}\,=\,-T_\textrm{\tiny D6}\,\int\,\dd^7\xi\,e^{-\Phi}\,{\rm Str}\left[\sqrt{-{\rm det}\left(g_7+\left(2\pi\alpha'\right)F\right)}\right] \,,
\end{equation}
where $g_7$ is the induced metric on the branes. In principle $F$ is a U($n_{\rm f}$) valued gauge field strength, but we will only turn on its SU($n_{\rm f}$) components.

The action (\ref{nADBI}) is an extremely complicated non-linear action for the gauge field. Fortunately, configurations that are selfdual (SD) or anti-selfdual (ASD) with respect to the appropriate metric minimize the action and therefore solve the equation of motion. Furthermore, preservation of supersymmetry selects either the SD or the ASD solution. Although we will be interested in SD configurations, for completeness we will present some of the formulas below for both cases. 

We will restrict ourselves to configurations in which the non-Abelian field strength is supported on the four-dimensional space inside the D6-branes that is orthogonal to the D2-branes, since such configurations carry the same type of D2-brane charge as the undissolved D2-branes. In this case, and assuming the (A)SD condition, it is easy to see that the determinant in the DBI action becomes a perfect square. As a consequence, the symmetrized trace reduces  to a normal trace over gauge indices and the action can be written as 
\begin{eqnarray}
S_\textrm{\tiny DBI}&=&-T_\textrm{\tiny D6}\,\int\,\dd^7\xi\,e^{-\Phi}\,\sqrt{-{\rm det}\,g_7}\,\,{\rm tr}\left(\mathbb{I}+\frac{\left(2\pi\alpha'\right)^2}{4}F^2\right)\nonumber\\[2mm]
&=&-T_\textrm{\tiny D6}\,\int\,e^{-\Phi}\,{\rm tr}\left(*_7\mathbb{I}+\frac{\left(2\pi\alpha'\right)^2}{2}\,e^{012}\wedge F\wedge*_4F\right)\,,
\end{eqnarray} 
where $e^{012}$ are vielbein along the (Minkowski) directions common to both the D2- and the D6-branes and $*_4$ is taken with respect to the remaining directions. We thus see that, for (A)SD configurations, all corrections to the Yang--Mills term contained in the DBI action vanish exactly. 

We will smear $\nf/n_{\rm f}$ of these stacks along the directions transverse to the color D2-branes in the same way as in Ref.~\cite{Faedo}. Note that we are distributing stacks of $n_{\rm f}$ overlapping flavor branes with a non-zero gauge field on them, and that the total number of D6-branes is $\nf$. The DBI action for the supersymmetric distribution of branes reads
\begin{equation}\label{nADBIsmeared}
S_\textrm{\tiny DBI}\,=\,-T_\textrm{\tiny D6}\,\int\,e^{-\Phi}\,\,{\rm tr}\left(\mathcal{K+I}\right)\wedge\Xi \,,
\end{equation}
where
\begin{equation}\label{square}
\mathcal{I}\,=\,\sigma\,\frac{\left(2\pi\alpha'\right)^2}{2}\,e^{012}\wedge F\wedge F
\,,\qquad\qquad
\sigma^2\,=\,1\,,
\end{equation}
and  $\mathcal{K}$ and $\Xi$ the usual calibration and smearing forms \cite{Faedo},  whose explicit expressions are  given below in \eqq{calibrationform} and \eqq{smearingform}, respectively.  We have already used (anti)selfduality, the sign $\sigma$ being $+1$ for SD configurations and $-1$ for ASD configurations.

The full D6-brane action is the sum of the DBI part plus the Wess-Zumino (WZ) term, which after smearing  takes the  form
\be
\label{WZ}
S_\textrm{\tiny WZ}=T_\textrm{\tiny D6}  \int {\rm tr}\left[C_7-\left(2\pi\alpha'\right)C_5\wedge F+\frac{\left(2\pi\alpha'\right)^2}{2}\,C_3\wedge F\wedge F
\right. 
-\left.\frac{\left(2\pi\alpha'\right)^3}{6}\,C_1\wedge F\wedge F\wedge F\right]\wedge\Xi \,.
\ee
We see that a non-zero $F$ induces several sources for the various Ramond--Ramond (RR) supergravity potentials. This leads to modifications of the Bianchi identities for their field strengths and, consequently, to modifications of the very definitions of these field strengths in terms of the potentials; for a detailed discussion see e.g.~Appendix A of \cite{Faedo2}. Taking these modifications into consideration and choosing to work with $F_2$ and $F_4$ as independent fields, the equations of motion for the supergravity RR fields are 
\begin{eqnarray}\label{RReoms}
\dd * F_4 +H \wedge  F_4 - \kappa^2T_\textrm{\tiny D6}\left(2\pi\alpha'\right)^2\, {\rm tr}\left( F\wedge F\right) \wedge \Xi &=&0\, , \nonumber\\[2mm]
\dd * F_2 +H \wedge * F_4 + \frac{ \kappa^2T_\textrm{\tiny D6}}{3}\left(2\pi\alpha'\right)^3\,{\rm tr}\left( F\wedge F\wedge F\right)\wedge\Xi&=&0 \,,
\end{eqnarray}
whereas the Bianchi identities read
\be\label{Bianchis}
\dd F_2 = -2 \kappa^2T_\textrm{\tiny D6}\, \Xi \,{\rm tr}\,\mathbb{I}=-2 \kappa^2T_\textrm{\tiny D6}\,n_{\rm f}\, \Xi \sac \dd F_4 = H \wedge F_2 \,.
\ee
If we were  considering an Abelian gauge field then the last equation would include a term proportional to $F \wedge \Xi$ that is here identically zero because SU($\nff$) matrices are traceless. We see from the equation of motion for $F_4$ that ${\rm tr}\left( F\wedge F\right)$ acts as a source for this field, consistent with the fact that an instanton density on the D6-branes corresponds to D2-brane charge dissolved on the D6-branes. This term therefore encodes the Higgsing of the gauge group and it will allow for its effective rank, as measured by the flux of $F_4$, to run with the holographic coordinate. 

The Neveu--Schwarz (NS) three-form $H$ only couples to the Abelian part of the branes' action, so its equation is unmodified with respect to the one considered in \cite{Faedo}. Therefore, as in that reference, we will set the NS form to zero, which solves identically its equation of motion, and ignore it hereafter. 

In contrast, the DBI action sources the other NS fields, namely the metric and  the dilaton, so it modifies their equations of motion. The equation for the dilaton is
\begin{equation}
R*1+4\dd*\dd\Phi-4\dd\Phi\wedge*\dd\Phi-\kappa^2\,T_\textrm{\tiny D6}\,e^\Phi\,{\rm tr}\left(\mathcal{K+I}\right)\wedge\Xi\,=\,0\,,
\end{equation}
whereas the Einstein's equations in the string frame take the form
\begin{equation}\label{D2D6Einstein}
R_{MN}+2\nabla_M\nabla_N\Phi\,=\,T_{MN}^{\rm IIA}+T_{MN}^{\rm sources}+T_{MN}^{\rm inst}\,,
\end{equation}
The explicit expressions for the stress tensors sourced by supergravity fields, $T_{MN}^{\rm IIA}$, and by the flavor branes $T_{MN}^{\rm sources}$, can be found in \cite{Faedo}. The contribution from the gauge field, $T_{MN}^{\rm inst}$, can be computed by taking into account that it only couples to the Minkowski part of the metric through its determinant $e^{012}$, as seen in (\ref{square}). This is in accordance with the results of \cite{Gibbons}, where it was shown that the energy-momentum tensor of (A)SD fields vanishes in non-linear electrodynamics, corresponding to the fact that the action can be reduced to $F\wedge F$, which is clearly topological.   

In the form written in (\ref{D2D6Einstein}), where the Ricci scalar has been eliminated using the dilaton equation of motion, this means that the transverse directions just pick a contribution
\begin{equation}
T_{mn}^{\rm inst}*1\,=\,g_{mn}\frac{\kappa^2\,T_\textrm{\tiny D6}}{2}\,e^\Phi\,{\rm tr}\,\mathcal{I}\wedge\Xi\,,
\end{equation}
with $m, n$ running along the radial and internal directions. Similarly for the Minkowski components we find:
\begin{equation}
T_{\mu\nu}^{\rm inst}*1\,=\,-g_{\mu\nu}\frac{\kappa^2\,T_\textrm{\tiny D6}}{2}\,e^\Phi\,{\rm tr}\,\mathcal{I}\wedge\Xi\,.
\end{equation}
This completes the set of equations to be solved for (A)SD configurations of the gauge field on the flavor branes. 

In order to solve the equations above we will first write an appropriate ansatz. Regarding the RR forms, we proceed by noting that  the only modification with respect to the situation with no instanton\footnote{Note that $F\wedge F\wedge F\wedge\Xi$  vanishes identically since neither $F$ nor $\Xi$ have components along the Minkowski directions.} is the instanton source for the four-form in the first equation in \eqq{RReoms}. Given that we expect this to encode color charge that will run with energy due to the Higgsing, we take the  ansatz
\begin{eqnarray}
F_2&=&\Qf (r) \,J \,,\nonumber\\[2mm]
F_6&=& * F_4 = \frac{\QQc (r)}{6}\,J\wedge J\wedge J\,,
\end{eqnarray}
where we allow  the effective  number of D2-branes, proportional to 
$\QQc (r)$, to depend on the radial coordinate.  $\Qf (r)$ is a constant if the quarks are massless but it becomes $r$-dependent for massive quarks, so in general we parametrize it as 
\be 
\label{dep}
\Qf (r) = \Qf \, p(r) \,,
\ee
with $\Qf$ a constant. The function $p(r)=1$ for massless quarks, whereas for massive quarks  $p(r)$ is a monotonically increasing function of $r$ with $p(r_m)=0$ and $p(\infty)=1$. The position $r_m>0$ at which $p(r)$ vanishes is the minimum distance between the D6-branes and the D2-branes and it is related to the quark mass through 
\be
m_{\rm q}=\frac{r_m}{2\pi\ell_s^2} \,.  
\label{mq}
\ee
The total D6-brane charge $\Qf$, which is proportional to the number of flavors in the gauge theory,   is given by 
\be
\label{see}
\Qf \, = \, \frac{(2\pi\ell_s)g_s}{V_2}\, \nf  \, = \,
 \frac{2\pi \ell_s^2}{V_2} \, \lambda \frac{\nf}{\nc} \,,
\ee
where the dimensionless volume $V_2=\int J$ and $\lambda$ is the 't Hooft's coupling $\lambda$, which in three dimensions  has dimensions of mass. This charge sets the energy scale below which the flavor degrees of freedom start to dominate the dynamics \cite{Faedo},  
\be\label{flavscale}
U_{\textrm{\tiny flavor}} \sim \frac{\Qf}{\ell_s^2} \sim \lambda\,\frac{\nf}{\nc}\,.
\ee

The fact that the globally-defined two-form $J$ is non-closed is crucial  to solve the Bianchi identity for $F_2$ in (\ref{Bianchis}). The exterior derivative of $J$, Eq.~\eqq{dif}, fixes the smearing form to be 
\begin{equation}
\label{smearingform}
\Xi\,=\,-\frac{\Qf}{2\kappa^2T_\textrm{\tiny D6}\,n_{\rm f}}\left[p'\,\dd r\wedge J+ 3\,p\,{\rm Im}\,\Omega\right]\,,
\end{equation}
where $'$ denotes differentiation  with respect to $r$.

In order to solve the equation of motion for $F_4$ we now need an ansatz for the instanton density. Since we wish to construct the  corresponding backreaction within the same ansatz as in \cite{Faedo}, the instanton density must not break any of the symmetries assumed in that reference. For this to be true the instanton density must be expressible in terms of the invariant forms of the original solution and it must therefore take the form 
\begin{equation}\label{psi}
{\rm tr}\left( F\wedge F\right)\wedge\Xi\,=\,\Psi(r)\,\dd r\wedge{\rm Re}\,\Omega\wedge\Xi\,,
\end{equation}
where $\Psi(r)$ is a function with dimensions of mass that we will solve for below.  The equation of motion for the four-form fixes the (derivative of the) D2-brane charge density to be 
\begin{equation}
\label{QCC}
\QQc' (r) \,=\,\frac{6\,\Qf (r)}{n_{\rm f}}\,\left(2\pi\alpha'\right)^2\,\Psi (r)\,.
\end{equation}
We thus see that $\Psi(r)$ directly encodes the running color charge. 
In particular, if  $\Psi=0$ then the  charge is just a constant  and we recover the solutions of \cite{Faedo}. 

Since by assumption the instanton density (\ref{psi}) breaks no symmetries of the original background we can use the same metric and dilaton ansatz as in the original solutions,
\begin{eqnarray}
\dd s_{\rm s}^2&=&h^{-\frac12}\,\dd x^2_{1,2}+h^{\frac12}\,e^{2\chi}\,\left[\dd r^2+r^2\,\dd s_6^2\left({\rm NK}\right)\right]\,,\nonumber\\[2mm]
e^{\Phi}&=&h^{\frac14}\,e^{3\chi}\,,
\end{eqnarray}
where $\dd s_6^2\left({\rm NK}\right)$ is the  the metric of a nearly K\"ahler manifold normalized to have a Ricci scalar $R_6=30$. The functions $h$ and $\chi$ depend only on the radial coordinate $r$ and were determined in \cite{Faedo} by solving  two first-order BPS equations.  In terms of these functions the calibration form that enters the DBI action takes the form 
\be
\mathcal{K}=h^{\frac14}\,e^{4\chi}\,\dd^3x\wedge  \left(
r^3\,\dd r\wedge{\rm Re}\,\Omega+\frac12\,r^4\,J\wedge J \right) \,.
\label{calibrationform}
\ee

Using the  ansatz above it is possible to check that, provided we consider SD configurations, corresponding to $\sigma=+1$ in (\ref{square}),  the only modification 
of the first-order BPS equations of \cite{Faedo} is the replacement $\Qc\to\QQc (r)$, so  they read
\begin{eqnarray}
\label{bps}
\chi'&=&\frac{\Qf (r)}{r^2}\,e^{2\chi}\,,\nonumber\\[4mm]
h'&=&-\frac{\QQc (r)}{r^6}\,e^{-2\chi}-\frac{3\,\Qf (r)}{r^2}\,e^{2\chi}\,h\,.
\end{eqnarray}
It is also easy to see that, with the sign choice $\sigma=+1$, there is an exact cancellation between the instanton contribution to the DBI action \eqq{nADBIsmeared} and the second term in the WZ action (\ref{WZ}),\footnote{Note that, with the convenient choice \eqq{vol},  $C_3=e^{-\Phi}\,e^{012}$ for $F_4=\dd C_3=*F_6$.}  as required by the no-force condition for supersymmetric configurations  of branes.

Since the equation for the function $\chi$ does not depend on the D2 charge, its solution is unmodified and reads
\begin{equation}
\label{pepe}
e^{-2\chi(r)}=1+2\int_r^\infty\,\Qf \, (z)\,\frac{\dd z}{z^2}\,,
\end{equation}
where we have fixed an integration constant by rescaling the radial and Minkowski coordinates. In contrast, 
the warp factor is determined by the color charge distribution through 
\begin{equation}\label{warpfactor}
h (r) \,=\,e^{-3\chi (r)}\,\int_r^\infty\,\,\frac{\QQc(y)\,e^{\chi(y)}}{y^6}\,\,\dd y\,,
\end{equation}
where we have again fixed an integration constant by taking the D2-brane decoupling limit. For constant $\QQc=\Qc$ the integral reduces to the one obtained in $\cite{Faedo}$. 

The only remaining task is to compute the instanton density $\Psi(r)$ associated to a selfdual field strength. Selfduality is a metric-dependent property, so the particular equations governing it are specific to each of the NK manifolds that we will consider in the next subsections.

\subsection{Super Yang--Mills theory}
\subsubsection{Massless quarks}
In this case the internal geometry is a six-dimensional sphere, whose metric  can be written as
\begin{equation}
\label{s6}
\dd \Omega_6^2\,=\dd\theta^2+\sin^2\theta\,\dd\Omega_3^2+\cos^2\theta\,\dd\Omega_2^2 \,,
\end{equation}
with $\dd \Omega_n^2$ the metric of ${\rm S}^n$. For massless quarks the D6-branes wrap equatorial three-spheres such as the one sitting at  $\theta=\pi/2$. 
The induced metric on any stack of branes is therefore 
\bea
\dd s_{7}^2 &=& h^{-\frac12}\,\dd x^2_{1,2}+h^{\frac12}\,e^{2\chi}\,\left(\dd r^2+r^2\,\dd \Omega_3^2\right) \nonumber \\[2mm]
&=& h^{-\frac12}\,\dd x^2_{1,2}+h^{\frac12}\,e^{2\chi}\,\dd z_4^2\,,
\eea
where $\dd z_4^2$ is the metric of $\mathbb{R}^4$. The instanton will be defined in this Euclidean space. Since the self-duality equations are conformally invariant,  we can ignore the warp factor $h$ and the function $\chi$. 

Our final goal is to construct SU($n_{\rm f}$) instantons whose instanton density is spherically symmetric so that symmetries are preserved. Since this can be done using SU(2) instantons as building blocks,  we will consider first this case in detail and then show how to generalize it. 

Let us write the metric of the three-sphere as
\begin{equation}
\label{Eq.S3metric}
\dd\Omega_3^2\,=\,\omega_1^2+\omega_2^2+\omega_3^2\,,
\end{equation}
with $\omega_i$ left-invariant forms satisfying 
\be
\dd\omega_i=\epsilon_{ijk}\,\omega_j\wedge\omega_k \,.
\label{normalized}
\ee 
We then take the following SU(2) gauge potential\footnote{A possible term $a_r(r)\,\dd r\otimes\mathbb{I}$ is pure gauge since it does not contribute to $F$.}
\begin{equation}\label{gaugepotential}
A\,=\,a_1(r)\,\omega_1\otimes\tau^1+a_2(r)\,\omega_2\otimes\tau^2+a_3(r)\,\omega_3\otimes\tau^3 \,,
\end{equation}
where $\tau^i$ are Hermitian generators obeying 
\be
\left[\tau^i,\tau^j\right]=i\epsilon^{ij}{}_k\tau^k \sac {\rm tr}\left(\tau^i\tau^j\right)=\delta^{ij}/2 \,. 
\ee
Note that the functions $a_i$ are dimensionless. 
The gauge potential is of course gauge-dependent, and we are choosing to lock the gauge index with the index of the SU(2) left-invariant forms. With this we compute the field strength 
\bea
F  &=&  \dd A+i\,A\wedge A \nonumber \\[2mm]
 &=& \Big[ a_1'\,\dd r\wedge\omega_1+\left(2a_1-a_2a_3\right)\omega_2\wedge\omega_3\Big]\otimes\tau^1+ \cdots \,,
\eea
where the dots stand for cyclic permutations of the indices $1, 2, 3$. The  gauge-invariant instanton density is then given by
\begin{equation}\label{instdensity}
 {\rm tr}\left( F\wedge F\right) \,=\,\frac{\dd}{\dd r}\Big[  \left(a_1^2+a_2^2+a_3^2\right)-\,a_1a_2a_3\Big]\,\dd r\wedge\omega_1\wedge\omega_2\wedge\omega_3\,.
\end{equation}
Requiring that $F$ be (A)SD with respect to the flat metric on $\mathbb{R}^4$,  
\begin{equation}
F=\sigma\,*_4F\,,\qquad\qquad\sigma^2=1 \,,
\end{equation}
yields the equation
\begin{equation}
a_1'\,=\,\frac{\sigma}{r}\left(2a_1-a_2\,a_3\right)
\end{equation}
and its two cyclic permutations.  The sign $\sigma$ is the same that appears in (\ref{square}). Notice that the non-linear terms are present because of the non-Abelian nature of the instanton and are crucial to obtain a regular solution. Given the rotational symmetry on S$^3$ of our problem, it is consistent to take all components to be equal, 
\be
a_1=a_2=a_3=a \,, 
\label{equal}
\ee
in which case the solution is 
\begin{equation}\label{single}
a\,=\,\frac{2\,r^{2\sigma}}{r^{2\sigma}+\Lambda^2}\,.
\end{equation}
The only free parameter is the integration constant $\Lambda$, which has dimensions of length. On the gravity side $\Lambda$ is the size of the instanton, whereas on the gauge theory side it corresponds to the vacuum expectation value (VEV) of one of the (s)quark bilinear operators that   parametrize  the Higgs branch \cite{Erdmenger:2005bj}: 
\be
v = \frac{\Lambda}{2\pi \ell_s^2}\,.
\ee
We will come back to this operator at the end of this section. For the moment it suffices to note that the gauge group is spontaneously broken at the energy scale set by $v$. 

As we already mentioned, the supersymmetric solution has a selfdual field strength so we will take $\sigma=+1$ henceforth. Furthermore, the instanton number is
\begin{equation}
k\,=\,\frac{1}{8\pi^2}\,\int\,{\rm tr}\left( F\wedge F\right)\,=\,12\,\Lambda^4\,\int_0^\infty\,\frac{r^{3}}{\left(r^{2}+\Lambda^2\right)^4}\,\dd r\,=\,1\,.
\end{equation}
This solution can be constructed on each stack of branes and then smeared together with them to match (\ref{psi}). The agreement is ensured by the fact that Re$\,\Omega$ is a calibration form for supersymmetric branes on NK manifolds \cite{Gutowski, Koerber}. In other words, the pullback of Re$\,\Omega$ together with the radial direction is the space in which we construct the instanton, so smearing it will give precisely (\ref{psi}). For the six-sphere, the pullback of Re$\,\Omega$ is the volume of the three-sphere wrapped by the D6-branes, as can be checked using the explicit form of the NK structure given in \cite{Faedo}. As a consequence, comparing with (\ref{instdensity}) we get that $\Psi(r)$ is related to the instanton potentials through 
\be
\label{thisone} 
\Psi (r) = \frac{\dd}{\dd r}\left[\left(a_1^2+a_2^2+a_3^2\right)-\,a_1\,a_2\,a_3\right]
= \frac{\dd}{\dd r}\left[3\,a^2-a^3\right]\,.
\ee

The fact that the solution above has instanton number 1 is not surprising given that we have required that the instanton density ${\rm tr}\left( F\wedge F\right)$ be spherically symmetric. More general SU(2) instanton configurations with $k>1$ break this symmetry. Intuitively, one can think of these configurations as built out of $k$ individual  instantons centered at different points, which is why they break  rotational symmetry. These configurations should certainly give rise to interesting supergravity solutions upon backreaction, but such solutions are outside the scope of our ansatz. See Appendix~\ref{appendixA}, however, for an approximate solution of this type. 

As we will now see, the limitation $k=1$ is not present for SU($n_{\rm f}$) instantons. In this case spherically symmetric configurations can be constructed by embedding the SU(2) solution above into SU($n_{\rm f}$). The different values of $k$ result from the different ways in which SU(2) can be embedded  into SU($n_{\rm f}$) (see \cite{Vandoren:2008xg} for a  review of this procedure). A trivial example consists of embedding the SU(2) solution as a $2\times2$ diagonal block of the $n_{\rm f}\times n_{\rm f}$ matrix, setting to zero the rest of the components of the SU($n_{\rm f}$) matrix. Less trivially, one can embed several SU(2) solutions as various $2\times2$ diagonal blocks inside the SU($n_{\rm f}$) matrix.  More generally, one may choose different irreducible representations of SU(2) with spins $j_n$ and use them to embed several SU(2) instantons as  
$(2j_n +1)\times (2j_n+1)$ diagonal blocks inside the SU($n_{\rm f}$) matrix. The only condition is that the combined size of these matrices fit inside an SU($n_{\rm f}$) matrix, namely that 
\be
\sum_{n} (2j_n+1) \leq \nff \,.
\ee
Note that each of the SU(2) instantons that we are using as a building block can have a different size $\Lambda_n$. Thus, for each of these instantons the gauge potential is still given by (\ref{gaugepotential}) and (\ref{single}) with $\Lambda$ replaced by 
$\Lambda_n$. In contrast, since for different representations the generators $\tau^i$ are normalized differently, one finds that the instanton charge of each of this building blocks is 
\begin{equation}
k_n \,=\,\frac23\,j_n\left(j_n+1\right)\left(2j_n+1\right)\,.
\end{equation}
The factor $j_n(j_n+1)$ comes form the quadratic Casimir, while $(2j_n+1)$ is the dimension of the representation. We thus arrive at the final conclusion that the total charge and the total instanton density are given by 
\be
k=\sum_n k_n \sac 
\Psi (r) = \sum_n k_n \Psi_n(r) \,,
\label{total}
\ee
with $\Psi_n(r)$ the instanton density of a $k=1$ SU(2) instanton of size 
$\Lambda_n$.  The fact that different building blocks can have different sizes is dual on the gauge theory to the statement that different operators can acquire different VEVs, thus breaking the gauge group at different scales $v_n$.  In most of what follows we will focus on the case of a single irreducible representation of SU(2), namely on the case  in which the sums above consist of only one term.

\begin{figure}[t!!]
\begin{center}
\includegraphics[width=0.60\textwidth]{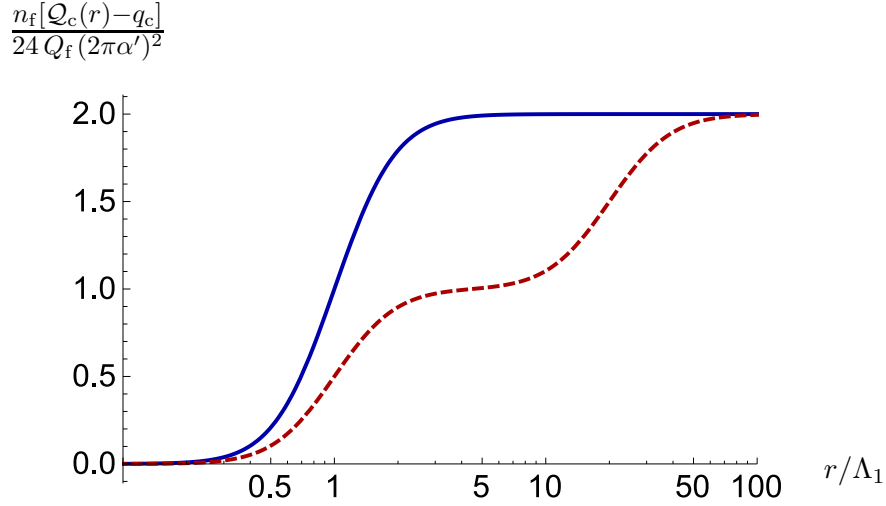}
\put(-290,175){\Large $\frac{n_{\rm f}\left[\QQc (r)-\qc\right]}{24\, \Qf\, (2\pi\alpha')^2}$}
\put(15,10){$ r/\Lambda_1$}
\caption{\small \label{plotted}
Integrated D2-brane charge densities carried by SU($\nff$) instantons built with two SU(2) instantons with $k_1=k_2=1$ and sizes such that 
$\Lambda_2/\Lambda_1=1$ (continuous, blue curve) and 
$\Lambda_2/\Lambda_1=20$ (dashed, red curve).}
\end{center}
\end{figure}

We now have all the ingredients to construct the running color charge. For an irreducible representation the density is simply
\be
\label{Psi}
\Psi (r) =  k\,\frac{\dd}{\dd r}\left[\left(a_1^2+a_2^2+a_3^2\right)-a_1a_2a_3\right] 
=k\,\frac{\dd}{\dd r}\left[3\,a^2-a^3\right]\,,
\ee
where in the last equation we have used rotational symmetry along the cycle wrapped by each stack of branes. Moreover, for massless quarks $\Qf$ is just a constant given by \eqq{see}, so replacing \eqq{Psi} in \eqq{QCC}, using \eqq{single} and  integrating we obtain 
\bea
\label{runcharge}
\QQc (r) &=& \qc+\frac{6\,\Qf\,k\,\left(2\pi\alpha'\right)^2}{n_{\rm f}}\,a^2\left(3-a\right) 
\nonumber\\[2mm]
&=& \qc+\frac{24\,\Qf\,k\,\left(2\pi\alpha'\right)^2r^4\left(r^2+3\Lambda^2\right)}{n_{\rm f}\,\left(r^2+\Lambda^2\right)^3}\,.
\eea
The  integration constant $\qc$ is physically the number of color branes that have not dissolved inside the flavor branes. The function $\QQc (r)- \qc$ is the integrated color charge carried by the instanton. In \fig{plotted} we plot this function  for SU($\nff$) instantons built with two SU(2) instantons with $k_1=k_2=1$ and sizes such that 
$\Lambda_2/\Lambda_1=1$ and $\Lambda_2/\Lambda_1=20$. Starting at large $r$, we see that in the first case the decrease in the gauge group takes place at a single scale $\Lambda_1=\Lambda_2$. In contrast, in the second example a partial breaking occurs at each of the two different scales $\Lambda_2$ and $\Lambda_1$, in between which the rank of the gauge group remains approximately constant.

We can now solve for the functions in the metric. Since for massless quarks $\Qf$ is a constant, Eq.~\eqq{pepe} can be simply integrated with the result
\be
e^{-2\chi (r)}\,=\,1+\frac{2\,\Qf}{r}\,. 
\ee
The warp factor $h$ can also be integrated in closed form, but the result is lengthy and not very illuminating. Its leading terms in the UV and IR can be inferred from the large- and small-$r$ behavior of the color charge. In the UV, when $r\to\infty$, the leading term is
\begin{equation}
h\,=\,\frac{\Qc}{5\,r^5}+\cdots 
\ee
with
\be
 \Qc=  \qc+\frac{24\,\Qf\,k\,\left(2\pi\alpha'\right)^2}{n_{\rm f}}\,.
 \label{wemay}
\end{equation}
This is the expected behavior for a configuration with total D2-charge $\Qc$, and we see that the second term on the right hand side is the contribution from the smeared instantons. The quantization condition for the color charge  is 
\be\label{Eq.chargequanti}
Q_c\,=\,\frac{(2\pi\ell_s)^{5}g_s}{V_6}\,\nc = 
\frac{(2\pi)^5 \ell_s^6}{V_6} \, \lambda  \,.
\ee
This, together with \eqq{see} and the relation between the volumes $12 V_6 = V_2V_3$, with $V_3$ the volume of the three-cycle wrapped by each stack of branes (in this case $V_3=2\pi^2$ for a three-sphere), means that we may rewrite Eq.~\eqq{wemay} as 
\be
\label{wemayindeed}
 \nc= \ncc + \frac{k \nf}{n_{\rm f}} \,,
\end{equation}
where $\ncc$ is the number of color branes that are not dissolved inside the flavor branes and $k \nf /n_{\rm f}$ is the number of color branes carried by the instantons. This is exactly the expected result, since we have $\nf /n_{\rm f}$ stacks of flavor branes and each of them carries an instanton with $k$ units of charge. 

We now turn to the IR behavior of the warp factor. The  $r\to0$ limit of the color charge  is $\QQc\to\qc$. If $\qc \neq 0$ then the leading term in the warp factor is 
\begin{equation}
h\,=\,\frac{4\,\qc\,\Qf}{9\,r^6}+\cdots\,.
\end{equation}
This is  the same behavior as in \cite{Faedo} except for the replacement of the total color charge $\Qc$ by the IR charge $\qc$. This is expected, since the charge carried by the instanton decouples in the far IR. In the gauge theory, this is exactly the physics of Higgsing: the gauge group is spontaneously broken, so its effective rank decreases with the energy scale. If $\qc \neq 0$ then the breaking is only partial and the unbroken gauge group in the far IR is SU($\ncc$). In this case the theory flows to a fixed point and the IR geometry is AdS$_4 \times \mbox{S}^6$ with a radius and a constant dilaton given by those of  \cite{Faedo} with the replacement $\nc \to \ncc$:
\bea
L(\ncc) &=&\frac{4\pi\ell_s}{3\sqrt{3}}\,\left(\frac{V_2}{V_6}\right)^\frac14\,\left(\frac{\ncc}{\nf}\right)^\frac14\,, \nonumber \\[2mm]
e^{\Phi (\ncc)}&=& \frac{1}{g_s}\,\frac{1}{2\sqrt{3}}\,\left(\frac{V_2^5}{V_6}\right)^\frac14\,\left(\frac{\ncc}{N_f^5}\right)^\frac14\,.
\eea
If $\qc=0$ then the leading IR behavior of the warp factor is 
\begin{equation}
h\,=\,\frac{288\,k\,\left(2\pi\alpha'\right)^2\Qf^2}{n_{\rm f}\,\Lambda^4\,r^2}+\cdots\,.
\label{leads}
\end{equation}
The corresponding solution is singular in the IR both in Einstein and string frames.

\begin{figure}[t!]
\begin{center}
\includegraphics[width=0.57\textwidth]{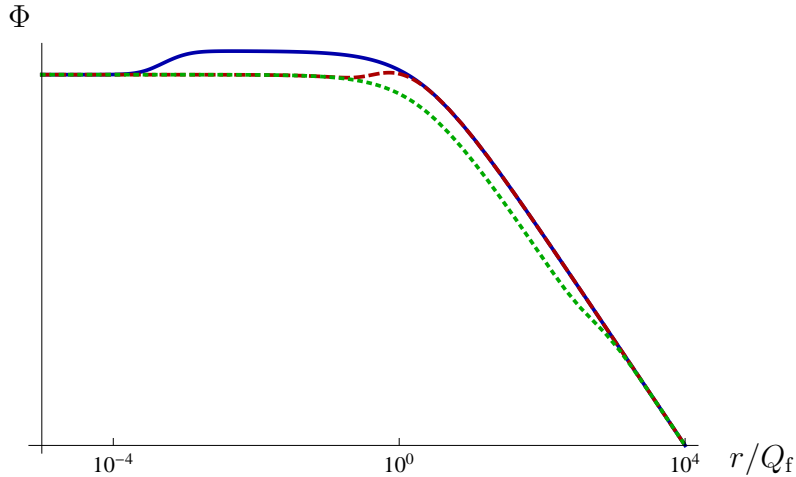}
\put(-260,170){\large $\Phi$}
\put(10,4){\large $r/\Qf$}
\caption{\small \label{Dilaton}
Dilaton as a function of the radial position, which is dual to the gauge theory energy scale. The values of $\Qc$ and $\Qf$ are the same for all the curves so that the asymptotic behaviors coincide. In contrast, the ratio $\Qf/\Lambda$ decreases from top to bottom. Specifically, the curves correspond to the ratios $10^3$ (blue, full line), $10^0$ (red, dashed line) and $10^{-3}$ (green, dotted line) respectively.}
\end{center}
\end{figure}

The Renormalization Group (RG) flows described by the solutions that we have just discussed fall into different qualitative classes depending on the hierarchies between  several energy scales (for this discussion we assume that $\qc \neq 0$). Consider first a single irreducible representation of SU(2). In this case the solution is completely characterized by  $v$ and $\uf$, which control the scale at which the Higgsing takes place and the scale at which the dynamics becomes approximately conformal, respectively.  If $v \gg \uf$ then, starting at high energies,  the RG flow first transitions at the scale $v$ from a SYM theory with $\nc$ colors to a SYM theory with $\ncc$ colors. At a lower scale $\uf$ the flow then enters 
 a conformal region. This situation is illustrated by the green, dotted curve in Fig.~\ref{Dilaton}, which shows the dilaton profile for this case. We see that at large $r$ there is first a transition between two curves with the same slopes, whereas at a smaller $r$ the dilaton becomes constant with a value $\Phi(\ncc)$, as appropriate to the AdS$_4$ solution with radius $L(\ncc)$.  

If instead $\uf \gg v$ then the RG flow first transitions at the scale $\uf$ from SYM with $\nc$ colors to a conformal theory described by the AdS$_4$ geometry with radius $L(\nc)$ and a constant dilaton $\Phi(\nc)$.  At the lower scale $v$ the Higgsing takes place and the theory evolves to a new conformal theory with smaller values of the radius and dilaton $L(\ncc)$ and $\Phi(\ncc)$, respectively. This is illustrated by the continuous, blue curve in Fig.~\ref{Dilaton}. 

Finally, if $v \sim \uf$, then there is a unique transition in which  the effective number of colors gets reduced at the same time that  the  dynamics becomes approximately conformal. This is illustrated by the dashed, red curve  
Fig.~\ref{Dilaton}.

Consider now the solution associated to a general SU($\nff$) instanton built with several SU(2) instantons with sizes $\Lambda_n$. For simplicity assume that all the corresponding scales $v_n$, as well as $\uf$, are hierarchically separated from one another. Then at each scale $v_n > \uf$ the theory transitions between two  SYM theories with different numbers of colors. In contrast, at each $v_n < \uf$ the theory transitions between two approximately conformal theories dual to two AdS$_4$ solutions with different radii and dilatons. Therefore the RG flow  exhibits quasi-conformal behavior, also known as ``walking dynamics'', in all the regions in between the scales $v_n < \uf$. The size of each walking region is  the difference between the two corresponding   $v_n$'s.  

We finally come back to the gauge theory operator dual to the instanton field. Upon dimensional reduction along the compact  directions, the components of this field give rise to scalar fields in the non-compact part of the geometry transforming in a specific representation of SU($\nff$). These scalar fields are then dual to scalar operators in the gauge theory transforming in the same representation of the global SU($\nff$) flavor symmetry of the gauge theory. In the case of the AdS$_4$ IR fixed point at the very end of the flow, these scalar operators have dimension $\Delta=13/3$ and 
the scalar fields in AdS$_4$ correspond to turning on a source for the  operators in the gauge theory.\footnote{There are also sources for operators of dimensions 6 and 11/3 that are present even in the absence of the instanton \cite{Faedo}.}  The simplest way to see this is to change to the AdS-adapted radial coordinate $\rho^2 = r^3/ \Qf$, so that the leading term for the gauge potential scales as $\mathcal{O}\left(\rho^{4/3}\right)$, giving $\Delta-3=4/3$.\footnote{Note that the flow cannot approach an IR fixed point with a non-zero vaccuum expectation value. Physically, this would lead to a breaking of the IR conformal symmetry. Mathematically, it would lead to a divergent term as $\rho\to 0$ of the form $\rho^{-\Delta}$.}  Only quadratic (or higher) combinations of these operators backreact  on  the dilaton and the warp factor, since these are only sensitive to SU($\nff$) gauge singlets. 

Determining the dimensions of the scalar operators in the UV is more subtle, since the dual gauge theory is not a CFT and consequently the geometry is not asymptotically AdS. However, this can be done by writing the solution in the so-called conformal frame and using the results from \cite{Kanitscheider:2008kd}. The outcome is as follows. The masses of the scalar fields fall in the region in which two quantizations are possible. In the so-called alternative quantization, the dual operators have dimension 1, as one would expect at weak coupling for an squark-bilinear operator, and the flow is triggered by a vacuum expectation value (VEV), as one would expect for a state on the Higgs branch.

\subsubsection{Massive quarks}

We will now consider the case of massive quarks. On the gravity side this is described by the fact that the D6-branes are now embedded non-trivially in the geometry sourced by the D2-branes. Specifically, the S$^3$ inside the S$^6$ wrapped by the D6-branes is no longer a constant equatorial sphere but becomes $r$-dependent. Writing the metric on the S$^6$ as in \eqq{s6}, the S$^3$ wrapped by the D6-branes is specified by a function $\theta(r)$. The induced metric then takes the form:
\begin{equation}\label{squaremassive}
\dd s_{7}^2\,=\,h^{-\frac12}\,\dd x^2_{1,2}+h^{\frac12}\,e^{2\chi}\,\Big[ \left(1+r^2\theta'^2\right)\dd r^2+r^2\sin^2\theta\,\dd\Omega_3^2\Big]\,.
\end{equation}
Crucially, the instanton field does not modify the equation for the embedding function $\theta(r)$ since, as we already mentioned, the stress energy tensor sourced by (A)SD fields vanishes. Thus we have as in \cite{Faedo} that a BPS embedding satisfies
\be\label{massembedd}
\cos\theta\,=\,\frac{r_m}{r}\,.
\ee
The radial position $r_m$ is related to the quark mass through \eqq{mq}. In the four-dimensional metric inside the squared brackets in \eqq{squaremassive}, which is no longer $\mathbb{R}^4$, the modified equation for (anti)selfduality reads:
\begin{equation}
a'\,=\,\frac{\sigma}{r}\,\frac{\left(1+r^2\theta'^2\right)^{1/2}}{\sin{\theta}}\,a\left(2-a\right)\,,
\end{equation}
where we have already assumed that all three components of the gauge potential are equal, as in \eqq{equal}. Using the embedding (\ref{massembedd}) this can be integrated straightforwardly with the result
\begin{equation}
a\,=\,\frac{2\left(r^2-r_m^2\right)^\sigma}{\left(r^2-r_m^2\right)^\sigma+\Lambda^2}\,.
\end{equation}
From this point onwards we will restrict ourselves again to the selfdual case, 
$\sigma=+1$.

A non-vanishing mass for the quarks  translates into a dependence of the flavor charge on the radial position as in \eqq{dep}. If all the quarks have the same mass then $p(r)$ is related to the embedding function through  \cite{Faedo}
\be
p(r)=\sin^4\theta (r)=\left[1-\left(\frac{r_m}{r}\right)^2\right]^2 \,\Theta\left(r-r_m\right)\,.
\ee
Inserting this in \eqq{QCC} we get the running color charge, which can be integrated analytically. The result is qualitatively the same as that  for massless quarks, although the precise form is more complicated due to the dependence of $\Qf(r)$ on the radial coordinate.  Notice that at $r=r_m$ the charge due to the instantons vanishes and we are left with $ \Qc=  \qc$ for $r<r_m$. 

The running color charge is continuous and differentiable (in fact $\mathcal{C}^3$). This ensures that the solution for the metric functions is $\mathcal{C}^2$ and reads
\begin{eqnarray}\label{massivesol}
e^{-2\chi}&=&\left\{\begin{array}{ll}
		1+\frac{16}{15}\,\frac{\Qf}{r_m}  & \qquad{\rm if } \,\,\, r<r_m \\[4mm]
		1+\frac{2\,\Qf}{r}\,\left[1-\frac23\left(\frac{r_m}{r}\right)^2+\frac15\left(\frac{r_m}{r}\right)^4\right] &\qquad{\rm if } \,\,\, r\ge r_m
	\end{array}
\right.\nonumber\\[4mm]
h&=&\left\{\begin{array}{ll}
		h_0+\left(1+\frac{16}{15}\,\frac{\Qf}{r_m}\right)\,\frac{\qc}{5\,r^5}  & \qquad{\rm if } \,\,\, r<r_m \\[4mm]
		e^{-3\chi}\,\int_{r}^\infty\frac{\QQc(y)}{y^6}e^{\chi}\dd y &\qquad{\rm if } \,\,\, r\ge r_m\,,
	\end{array}
\right.
\end{eqnarray}
with the constant $h_0$ fixed by continuity. 

As in the case of massless quarks, the solutions fall into several qualitative classes depending on the energy scales involved. For simplicity let us discuss here only the case of a single irreducible representation of SU(2). This solution is characterized by three physical scales. The scale $\uf$ (equivalently, $\Qf$) at which the quarks start to dominate the dynamics, the scale $\mq$  (equivalently, $r_m$) at which the quarks decouple from the dynamics, and the scale $v$ (equivalently, $\Lambda$) at which the Higgsing of the gauge group takes place. Depending on the hierarchy between these scales the physics is as follows. For this discussion we will assume that all the scales are well separated. If some of them coincide then some of the transitions below take place simultaneously. 

Suppose first that $\mq \ll v \ll \uf$. In this case the theory starts as SYM with $\nc$ colors in the far UV, it then enters the conformal region at $\uf$, it subsequently undergoes the Higgsing at $v$ and the consequent transition between two AdS spaces, and finally at the scale $\mq$ the quarks decouple. If $\qc\neq 0$ then at this point the theory enters again a pure SYM phase (i.e.~with no flavors) with $\ncc$ colors described by a D2-brane geometry. If instead $\qc=0$ then the gauge group is completely Higgsed and there are no degrees of freedom below $\mq$, i.e.~the theory is gapped. The geometry for $r<r_m$ is simply flat space and all the fluxes vanish. 

Suppose now that $\mq \ll \uf \ll v$. In this case the first transition is the Higgsing of the gauge group from SYM with $\nc$ colors to SYM with $\ncc$ colors, and it takes place at the scale $v$. The theory then enters the conformal phase at $\uf$ and finally at $\mq$ the quarks decouple. At this point the physics is as in the previous paragraph. 

Finally, if $\uf \ll \mq \ll v$, then the first transition at $v$ is again between two SYM theories, and at the second transition at the scale $\mq$ the quarks decouple. Therefore the theory never enters a conformal regime. Below the scale $\mq$ the physics is again as in the two previous cases. 

\subsection{Chern--Simons-Matter theory}

Given that the AdS geometry describing the IR fixed point in the absence of the instanton is a solution by itself \cite{Faedo}, we may detach it from the D2-brane part of the geometry and construct the following flows between two fixed points: 
\begin{eqnarray}
e^{-2\chi}&=&\frac{2\,\Qf}{r}\,,\\[2mm]
h&=&\frac{k\,\left(2\pi\alpha'\right)^2\Qf^2}{18\,n_{\rm f}}\, \Bigg\{
\frac{8\,n_{\rm f}\,\qc}{k\,\left(2\pi\alpha'\right)^2\Qf \,r^6}+\frac{5184}{r^2\Lambda^4}+\frac{432}{\Lambda^2\left( r^2+\Lambda^2\right)^2}+
\frac{1836}{\Lambda^4\left(r^2+\Lambda^2\right)} \nonumber \\[2mm]
&-&\frac{1755\sqrt{2}}{r^{3/2}\Lambda^{9/2}} \Bigg[ 
\pi+\operatorname{arcoth}\left(\frac{r+\Lambda}{\sqrt{2r\Lambda}}\right)+\arctan\left(1-\sqrt{\frac{2r}{\Lambda}}\right) + \arctan\left(1+\sqrt{\frac{2r}{\Lambda}}\right)
\Bigg] \Bigg\}
\nonumber
\end{eqnarray} 
This solution interpolates between two AdS geometries with radii related through 
\begin{equation}
\frac{L_{\textrm{\tiny UV}}}{L_{\textrm{\tiny IR}}}\,=\,
\frac{L(\nc)}{L(\ncc)} \,=\, 
\left( 1+\frac{k\,\nf}{n_{\rm f}\,\ncc} \right)^{1/4}>1\,,
\end{equation}
as corresponds to a change in the rank of the  gauge group from $\nc= \ncc +k\, \nf /n_{\rm f}$ to $\ncc$. The dual field  theory both in the UV and in the IR is a Chern--Simons-Matter theory \cite{Faedo}. As in the previous section, in the UV the masses of the scalar fields arising from the dimensional reduction of the instanton  fall in the region in which two quantizations are possible. In the alternative quantization, the flow is triggered in the UV conformal theory purely by the VEV of a dimension $\Delta_{\textrm{\tiny UV}}=4/3$ operator that induces the Higgsing. The IR fixed point is reached through sources for operators of dimensions $\Delta_{\textrm{\tiny IR}}=13/3,\,6$. The first one is the operator dual to the instanton, whereas the second one is an additional operator that gets sourced along the flow.

\subsection{Quiver-like theories}

According to the results in \cite{Faedo} there is a solution preserving $\mathcal{N}=1$ for every NK internal manifold. If the internal manifold is not the six-sphere, then the dual gauge model is presumably a quiver-like theory. Leaving aside the six-sphere, only three other regular examples are known: the product ${\rm S}^3\times{\rm S}^3$, the complex projective space $\mathbb{CP}^3$ and the flag manifold $F(1,2)\simeq{\rm SU(3)}/{\rm U(1)}^2$. We will only discuss the construction of the Higgs branch in the first two geometries, since the third case is rather similar to the second one and we do not expect new  features to arise.

\subsubsection{${\rm S}^3\times{\rm S}^3$ internal geometry}

The NK metric of ${\rm S}^3\times{\rm S}^3$, written in terms of two sets of left-invariant forms $\rho_i$ and $\eta_i$, normalized as above, reads
\begin{equation}
\dd s_6^2\,=\,\frac49\,\sum_{i=1}^3\left(\rho_i^2+\eta_i^2-\rho_i\,\eta_i\right)\,.
\end{equation}
There are only two inequivalent choices for BPS-embedded D6-branes. Either they wrap one of the two three-spheres, in which case
\begin{equation}
\rho_i\,=\,0,\,\,\eta_i\,=\,\omega_i\qquad\qquad{\rm or}\qquad\qquad\eta_i\,=\,0,\,\,\rho_i\,=\,\omega_i\,,
\end{equation}
or they wrap a diagonal combination of the two three-spheres, so that
\begin{equation}
\rho_i\,=\,\eta_i\,=\,\omega_i\,.
\end{equation}
In either case the induced metric on the D6-branes is
\begin{equation}
\dd s_{7}^2\,=\,h^{-\frac12}\,\dd x^2_{1,2}+h^{\frac12}\,e^{2\chi}\,\left(\dd r^2+\frac49\,r^2\,\dd \Omega_3^2\right)\,.
\end{equation}
We have to construct an (anti)selfdual solution with respect to the metric in parenthesis, which is not simply flat space but a cone.\footnote{Note, however, that the full induced metric is non-singular but $\mbox{AdS}_4 \times \mbox{S}^3$ in the IR, since the spacetime metric is $\mbox{AdS}_4 \times \mbox{S}^3 \times \mbox{S}^3$ \cite{Faedo}.} The very same ansatz (\ref{gaugepotential}) for the gauge potential can be adopted, and requiring (anti)selfduality leads to the following equation
\begin{equation}
a'\,=\,\frac{3\,\sigma}{2\,r}\,a\,\left(2-a\right)\,,
\end{equation}
where we have already assumed that all three components of the gauge potential are equal, as in \eqq{equal}. The regular solution is
\begin{equation}
a\,=\,\frac{2\,r^{3\sigma}}{r^{3\sigma}+\Lambda^3}\,,
\end{equation}
which is very similar to the one in flat space but with a different power of the radial coordinate. Supersymmetry requires again $\sigma=+1$. On the other hand, using an explicit construction of the NK structure, it can be seen that 
\begin{equation}
\rm{Re}\,\Omega|_{\textrm{\tiny D6}}\,=\,\left(\frac{2}{3}\right)^3\,\omega_1\wedge\omega_2\wedge\omega_3\,.
\end{equation}
Comparing with (\ref{psi}) and (\ref{instdensity}) gives
\begin{equation}\label{Eq.psiS3S3}
\Psi\,=\,k\,\left(\frac{3}{2}\right)^3\,\frac{\dd}{\dd r}\left[3a^2-a^3\right]\,=\,\frac{243\,k\,r^{5}\,\Lambda^6}{\left(r^{3}+\Lambda^3\right)^4}\,,
\end{equation}
from which it is straightforward  to get the running color charge. The instanton number, obtained by integrating the instanton density over the cone, is still $k$. 

It is also possible to solve for the warp factor according to (\ref{warpfactor}). The structure of the solution is very similar to that for the six-sphere. The UV geometry is again that of  $\ncc+k \nf/n_{\rm f}$ D2-branes. The IR is AdS if $\qc \neq 0$ or a singular geometry with $h\sim r^{-3/2}$ as $r \to 0$ if  $\qc=0$.  In the first case 
 the IR  deformation due to the instanton corresponds now to an operator of dimension $\Delta=5$, since the instanton field \eqq{Eq.psiS3S3} scales with a different power of the radial coordinate compared to the previous section.  If $\qc\ne0$ and $v < \uf$ then there is an energy range where the model is quasi-conformal. The D2-brane region in the UV can also be decoupled from the geometry to obtain a flow between two CFTs due purely to Higgsing with a VEV of a dimension $\Delta=2$ operator (in the alternative quantization).

\subsubsection{$\mathbb{CP}^3$ internal geometry}

We can also consider the complex projective space $\mathbb{CP}^3$ as a nearly NK  manifold, whose metric is squashed with respect to the Fubini--Study one. In this case the BPS branes wrap a squashed $\mathbb{RP}^3\simeq{\rm S}^3/\mathbb{Z}_2$ whose metric, after some manipulations as seen in \cite{Conde}, reads
\begin{equation}
\dd s_3^2\,=\,\frac12\left(\dd \theta^2+\sin^2\theta\,\dd \varphi^2\right)+\frac14\left(\dd\hat{\psi}+\cos\theta\,\dd\varphi\right)^2\,,
\end{equation}
with the ranges $0\le\theta<\pi$, $0\le\varphi<2\pi$ and $0\le\hat{\psi}<2\pi$. In particular, the range of $\hat{\psi}$ is not the one for a sphere. In terms of left-invariant forms, the induced metric on the D6-branes is thus
\begin{equation}
\dd s_{7}^2\,=\,h^{-\frac12}\,\dd x^2_{1,2}+h^{\frac12}\,e^{2\chi}\,\Big[
\dd r^2+\,r^2\,\left(2\left(\omega_1^2+\omega_2^2\right)+ \omega_3^2\right)
\Big]\,.
\end{equation}
Notice the squashing of the third direction with respect to the others, inherited from the squashing of the $\mathbb{RP}^3$ inside $\mathbb{CP}^3$. Once again, the ansatz (\ref{gaugepotential}) is suited for constructing selfdual solutions. The relevant set of equations is now
\begin{eqnarray}\label{cannot}
a_1'&=&\frac{\sigma}{r}\left(2a_1-a_2\,a_3\right)\,,\nonumber\\[2mm]
a_2'&=&\frac{\sigma}{r}\left(2a_2-a_3\,a_1\right)\,,\\[2mm]
a_3'&=&\frac{\sigma}{2\,r}\left(2a_3-a_1\,a_2\right)\,.\nonumber
\end{eqnarray}
The main new feature is that, due to the squashing, which breaks part of the rotational symmetry, we can set $a_1=a_2$ but $a_3\ne a_1$. These feature will also appear when we consider four-dimensional gauge theories in the next section. 

\begin{figure}[t!]
\begin{center}
\includegraphics[width=0.57\textwidth]{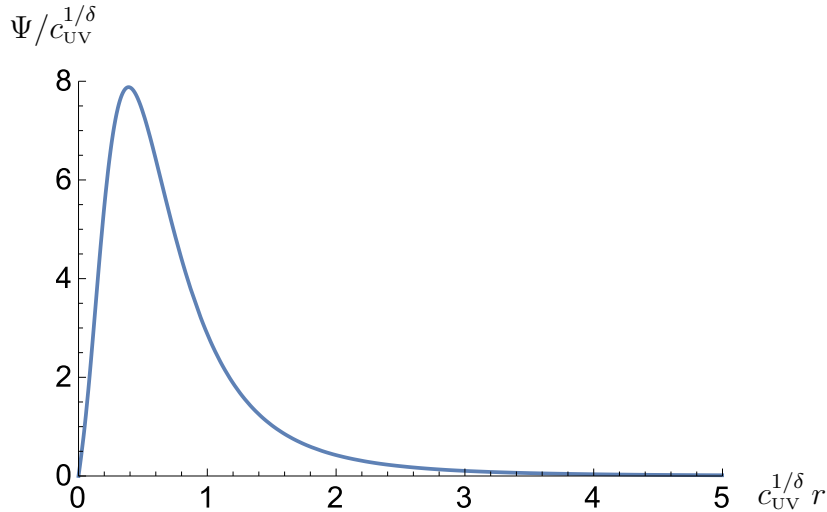}
\put(-270,180){\large $\Psi/c_\textrm{\tiny UV}^{1/\delta}$}
\put(10,4){\large $c_\textrm{\tiny UV}^{1/\delta} \, r $}
\caption{\small \label{instantonCP3}
Instanton density as a function of the radial position for the $\mathbb{CP}^3$ internal geometry. 
}
\end{center}
\end{figure}

The equations \eqq{cannot} cannot be integrated in closed form, but  it is possible to find a regular expansion around the IR, 
\begin{eqnarray}
a_1&=&c_1\left[r^2-\,c_3\,r^3+ \frac{c_3^2}{2} \,r^4+\mathcal{O}\left(r^5\right)\right]\,,\nonumber\\[2mm]
a_3&=&c_3\,r-\frac{c_1^2}{6}\,r^4+\mathcal{O}\left(r^5\right)\,,
\end{eqnarray}
and a normalizable solution around the UV,
\begin{eqnarray}
a_1&=&2-c_{\textrm{\tiny UV}}\,r^\delta+\frac{32+\sqrt{17}}{76}c_{\textrm{\tiny UV}}^2\,r^{2\delta}+\mathcal{O}\left(r^{3\delta}\right)\,,\nonumber\\[2mm]
a_3&=&2-\frac14\left(\sqrt{17}-1\right)c_{\textrm{\tiny UV}}\,r^{\delta}+\frac{1+3\sqrt{17}}{38}c_{\textrm{\tiny UV}}^2\,r^{2\delta}+\mathcal{O}\left(r^{3\delta}\right)\,,
\end{eqnarray}
where  
\be
\delta=\left(1-\sqrt{17}\right)/2<0 \,. 
\ee
Numerical integration determines  the IR constants $c_1, c_3$  in terms of the UV constant $c_\textrm{\tiny UV}$ as 
\be
\left( c_\textrm{\tiny UV} \right)^{1/\delta} \simeq 0.420884\,  c_3 \simeq 0.323277 \, c_1^{1/2} \,.
\ee
Note that the functional form of these relations is fixed by dimensional analysis, with the numerical integration fixing only the coefficients. The full solution, which is shown in Fig.~\ref{instantonCP3}, is therefore determined in terms of a single parameter with dimensions of length, $c_\textrm{\tiny UV}^{-1/\delta}$, which plays the role of the size of the instanton. 

From the NK structure we can read $\rm{Re}\,\Omega|_{\textrm{\tiny D6}}=2\,\omega_1\wedge\omega_2\wedge\omega_3$, and from this the running charge and the warp factor. The number of D2-branes induced by the instanton is $kN_f/2n_{\rm f}$, consistently with the fact that due to the orbifolding (as reflected in the range of $\hat{\psi}$) the instanton number is $k/2$. For this quiver, and due to the squashing that breaks rotational symmetry, different components of the gauge potential contain different powers of the radial coordinate and therefore are dual to operators of different dimensions. Specifically, in the IR we have  $\Delta_1=13/3$ for $a_1$ and $\Delta_2=11/3$ for $a_3$. There is nevertheless a single scale, corresponding to the size of the instanton, that determines the sources for both operators. 
 
Given that addition of flavor to the ABJM geometry produces a squashing of the $\mathbb{CP}^3$ analogous to the one we have discussed \cite{Conde}, the very same construction of instantons can be made for ABJM in the presence of backreacting flavor. In the solutions of \cite{Conde}, the squashing depends on the ratio between the number of flavors and the Chern--Simons (CS) level and it will appear both in the induced metric and the selfduality equations. As a consequence,  the radial powers of the instanton solution and the dimension of the dual operator  will depend on this ratio. The solutions constructed in this way will correspond to a flow between ABJM with flavor, which is conformal, and another conformal fixed point in the IR. The flow is triggered by an operator getting a VEV and the rank of the dual gauge group is reduced along the flow due to Higgsing.

\section{Four-dimensional gauge theories}\label{sec.4dgaugetheories}

We now turn our attention to $\mathcal{N}=1$ four-dimensional gauge theories. For simplicity we will only consider massless quarks. We will follow \cite{Benini} with slight differences in conventions and, as in that reference, work in Einstein frame, as opposed to the previous section where we worked in string frame. This means that some dilaton  factors must be taken into consideration when comparing the two sections. Otherwise  most of the setup is common  between the two cases, so we will give fewer details here.

The basic flavored solution was constructed in \cite{Benini} and corresponds to the backreaction of smeared D7-branes in the background of color D3-branes. The gravity solution possesses a singularity in the UV, which is dual to the presence of  a Landau pole in the gauge theory. In the IR the geometry approaches a spacetime that one may call a ``logarithmically-corrected AdS''. Despite the fact that the metric is AdS up to logarithmic terms, the geometry is singular. 

 The five-dimensional internal geometries of the flavored solutions are Sasaki--Einstein (SE) spaces. Their structure is that of a U(1) fibration over a K\"ahler--Einstein (KE) base. These manifolds are equipped with two real, globally defined forms, a one-form $\eta$ and a two-form $J$ (the K\"ahler form of the base),  satisfying the conditions
\be
\dd \eta = 2\,J \ , \qquad \qquad J \wedge J \wedge \eta = -2 \omega_5
\label{ext}
\ee
with $\omega_5$ the volume form of the five-dimensional  SE manifold.

We consider $\nf/\nff$ stacks of D7-branes, each of them consisting of $\nff$ overlapping D7-branes so that the total number of flavor branes is $\nf$. All the branes wrap the U(1) fiber generated by $\eta$,  but each stack wraps a different two-dimensional submanifold of the KE base. Through the smearing procedure, all these two-dimensional submanifolds are related to one another by the action of an isometry of the KE base. The DBI action for the smeared stacks of flavor branes  with (A)SD gauge fields on them can be written as in Eq.~\eqref{nADBIsmeared}, 
\begin{equation}
S_\textrm{\tiny DBI}\,=\,-T_{\textrm{\tiny D7}}\,\int\,{\rm tr}\left(e^{\Phi}\mathcal{K+I}\right)\wedge\Xi\,,
\end{equation}
where $\mathcal{K}$ and $\Xi$ are the usual calibration and smearing forms \cite{Benini},  whose explicit expressions are given below in \eqq{calibrationform4} and \eqq{smearingform4}, respectively. The instanton contribution is
\begin{equation}
\mathcal{I}\,=\,\sigma\frac{\left(2\pi\alpha'\right)^2}{2}\,e^{0123}\wedge F\wedge F\,,\qquad\qquad\qquad\qquad\sigma^2\,=\,1\,,
\end{equation}
with $e^{0123}$ vielbeins in the Minkowski directions. Notice in particular that the instanton term does not source the dilaton, in agreement with the fact that it represents dissolved D3-branes. Analogously the WZ term reads
\begin{eqnarray}
S_\textrm{\tiny WZ}&=&T_\textrm{\tiny D7}\,\int\,{\rm tr}\left[C_8-\left(2\pi\alpha'\right)C_6\wedge F+\frac{\left(2\pi\alpha'\right)^2}{2}\,C_4\wedge F\wedge F-\frac{\left(2\pi\alpha'\right)^3}{3!}\,C_2\wedge F\wedge F\wedge F\right.\nonumber\\[2mm]
&&\qquad\qquad\qquad\left.+\frac{\left(2\pi\alpha'\right)^4}{4!}\,C_0\,F\wedge F\wedge F\wedge F\right]\wedge\Xi\,.
\end{eqnarray}
This term alters several of the equations of motion and Bianchi identities for the RR forms.  In the case of interest here the only modifications are:
\begin{eqnarray}
\dd F_1&=&-2\kappa^2T_\textrm{\tiny D7}\,\Xi\,{\rm tr}\,\mathbb{I}=-2\kappa^2T_\textrm{\tiny D7}\,n_{\rm f}\,\Xi \ , \nonumber\\[2mm] 
\dd F_5&=&-  \kappa^2T_\textrm{\tiny D7}\,\left(2\pi\alpha'\right)^2\, {\rm tr}\left( F\wedge F\right) \wedge \Xi\,.
\label{f1}
\end{eqnarray}
Recall that the equation of motion for the five-form is the same as its Bianchi identity, so again we see that the instanton acts as a source of color charge. Einstein's equations are also modified with respect to the ones solved in  \cite{Benini} due to the presence of the instanton, and since we are working directly in Einstein frame, we have to add the contribution of the (anti)selfdual Yang--Mills field only along the Minkowski directions
\begin{equation}
T_{\mu\nu}^{\rm inst}\,*1\,=\,-g_{\mu\nu}\,\kappa^2\,T_\textrm{\tiny D7}\,{\rm tr}\,\mathcal{I}\wedge\Xi.
\end{equation}

Let us now present our ansatz. The SE internal manifold is a U(1) fibration over a KE base. The backreaction of the D7-branes causes a relative squashing between them \cite{Benini}, so the metric takes the form
\begin{equation}
\dd s^2\,=\,h^{-\frac12}\dd x_{1,3}^2+h^{\frac12}\,e^{2f}\left(\dd\rho^2+e^{2g-2f}\dd s_4^2+\eta^2\right)\,,
\end{equation}
with $\dd s_4^2$ the metric of the KE base, $\eta$ the fiber over it and $\rho$ the holographic radial coordinate. We recover the SE structure normalized to have curvature $R_5=20$ when $f=g$. The radial coordinate $\rho$ that we will work with in this section is dimensionless, whereas the warp factor $h$ has dimensions of (length)$^4$. This is a convenient choice here despite the fact that it implies  the somewhat unusual feature that the Minkowski coordinates have dimensions of  (length)$^2$. In terms of the functions above the calibration form that enters the DBI action takes the form 
\be
{ \cal K } = e^{ 2f+2g } \,  \dd^4 x \wedge \left( \dd \rho \wedge \eta \wedge J + 
\frac{1}{2} e^{2g-2f} \, J \wedge J \right) \,.
\label{calibrationform4}
\ee

Following the same logic as in the three-dimensional case, the only modification we expect is a varying  number of colors. Therefore we write the RR forms as
\begin{equation}
\begin{array}{rclcrcl}
F_5&=&\left(1+*\right)\,\QQc(\rho)\,\omega_5 \ , \\
F_1&=&\Qf(\rho)\,\eta \ ,
\end{array}
\end{equation}
where we have allowed for the number of D3-branes to run with the radial coordinate. As in the previous section $\Qf (\rho) =\Qf \, p(\rho)$, with $\Qf$ a constant, $p(\rho)=1$ for massless quarks and $p(\rho)$ a non-constant function of $\rho$ for massive quarks \cite{Bigazzi}. The total D7-brane charge is given by
\be
\Qf = \frac{1}{2\pi V_1} \lambda \frac{\nf}{\nc}= 
\frac{V_3}{8\pi V_5} \lambda \frac{\nf}{\nc} \,,
\label{D7quant}
\ee
with $\lambda=2\pi g_s \nc$ the 't Hooft coupling and $V_1=\int \eta$.  
Note that, unlike in the three-dimensional case,  now $\Qf$ and the 't Hooft coupling are dimensionless. 

The fact that the globally-defined one-form $\eta$ is non-closed is crucial  to solve the Bianchi identity for $F_1$ in (\ref{f1}). The exterior derivative of $\eta$, Eq.~\eqq{ext}, fixes the smearing form to be given by
\begin{equation}
\label{smearingform4}
\Xi = - \frac{\Qf}{2 \kappa^2 T_\mt{D7}  \nff}  \left[ p' \, \dd \rho \wedge \eta + 2 p\, J \right] \,,
\end{equation}
where  $'$ denotes differentiation  with respect to $\rho$. As stated above, we will only consider the massless case, so henceforth we will set $p=1$.

The three-dimensional submanifold wrapped by the calibrated branes has support along $J\wedge\eta$ inside the SE manifold, suggesting that the instanton density will take the form
\begin{equation}
{\rm tr}\left( F\wedge F\right)\wedge\Xi\,=\,\Psi(\rho)\,\dd \rho\wedge J\wedge\eta\wedge\Xi\,.
\end{equation} 
The equation of motion for the five-form then relates the running D3-charge  to the instanton density as
\begin{equation}\label{colorcharge}
\QQc' (\rho) \,=\,\frac{2\,\Qf}{n_{\rm f}} \,\left(2\pi\alpha'\right)^2\,\Psi (\rho)\,.
\end{equation}
Note that, unlike in the D2-D6 case, here not just both the gauge potentials $a_i$ but also  the instanton density $\Psi(\rho)$ are dimensionless.

Since the instanton density is related to the non-abelian gauge field exactly as in Eq.~\eqref{Psi} we have 
\be\label{D3charge}
\QQc = \qc + \frac{2\,k\,\Qf\,\left(2\pi\alpha'\right)^2}{n_{\rm f}} \, \left[  \left( a_1^2 + a_2^2 + a_3^2 \right) - a_1\, a_2\, a_3 \right]  \,,
\ee
with $\qc$ the number of D3-branes not dissolved inside the flavor branes. This equation is analogous to that in the first line of \eqq{runcharge} except that we have not assumed that all the $a_i$ are equal.

The BPS equations that determine the solution are the same as those in \cite{Benini} with the replacement of $\Qc$ by $\QQc$:
\begin{eqnarray}
\Phi'&=&\Qf\,e^{\Phi}\nonumber\\[2mm]
g'&=&e^{2f-2g}\nonumber\\[2mm]
f'&=&3-2\,e^{2f-2g}-\frac{\Qf}{2}\,e^\Phi\nonumber\\[2mm]
h'&=&-\QQc\,e^{-4g}\,.
\label{three}
\end{eqnarray}
For massless quarks $\Qf$ is a constant and the solution of the first three equations, which are independent of the color charge, is \cite{Benini}
\begin{eqnarray}\label{D3D7smeared}
e^{\Phi}&=&-\frac{1}{\Qf\,\rho}\,,\nonumber\\[2mm]
e^{2f}&=&-6\rho\,\left(1-6\rho\right)^{-2/3}\,e^{2\rho}\,,\nonumber\\[2mm]
e^{2g}&=&\left(1-6\rho\right)^{1/3}\,e^{2\rho}\,,
\end{eqnarray}
where we have already fixed some integration constants. The range of the radial coordinate is $\rho\in\left(-\infty,\,0\right)$. In the UV, corresponding to $\rho=0$, there is a singularity, as can be seen from the fact that the dilaton diverges. In the dual gauge theory this corresponds to the presence of a Landau pole.

\subsection{Super Yang--Mills theory}

When the dual gauge theory is SYM the SE manifold is the five-sphere. If in addition the quarks are massless, then each stack of D7-branes wraps an equatorial  three-sphere inside it. Taking the fiber to be the third direction in S$^3$, the induced metric reads
\begin{equation}
\dd s^2_8\,=\,h^{-1/2}\dd x_{1,3}^2+h^{1/2}\,e^{2f}\left[
\dd\rho^2+e^{2g-2f}\left(\omega_1^2+\omega_2^2\right)+\omega_3^2\right]\,,
\label{this}
\end{equation}
with $\omega_i$ the usual SU(2) left-invariant  one-forms. 

As above, we begin with the SU(2) instanton and we use it to construct an SU($\nff$) instanton. Adopting the same ansatz for the gauge potential as in Eq.~(\ref{gaugepotential}) with $r$ replaced by $\rho$, (anti)selfduality with respect to the metric inside the squared brackets in \eqq{this} results in the following equations:
\begin{eqnarray}
a_1'&=&\sigma\left(2a_1-a_2\,a_3\right) \,, \nonumber\\[2mm]
a_2'&=&\sigma\left(2a_2-a_3\,a_1\right)  \,, \nonumber\\[2mm]
a_3'&=&\sigma\,e^{2f-2g}\,\left(2a_3-a_1\,a_2\right)\,.
\end{eqnarray}
In the present conventions, the selfdual configuration solves the equations of motion and is BPS, so we restrict ourselves to $\sigma=1$. Because of the squashing of the metric we cannot take all the $a_i$ equal to one another but only $a_2=a_1$, in which case the equations reduce to
\begin{eqnarray}
a_1'&=& \xi \, a_1 \left(2- a_3\right)  \,,  \nonumber\\[2mm]
a_3'&=&e^{2f-2g}\,\left(2a_3-a_1^2\right)\,.
\end{eqnarray}
In the present situation $\xi=1$, but we will give some expressions below for generic $\xi$ because in Sec.~\ref{sec.otherquivers4d} we will be interested in the values $\xi=3/2$ and $\xi=3/4$.

We have not been able to solve the equations above analytically, but asymptotic solutions  can be found both in the UV and in the IR. In the UV we have 
\begin{eqnarray}\label{eq.UVexpansionD3D7}
a_1&=&c_1\left[1 - \xi \left(c_3-2\right)\,\rho+\frac{\xi^2}{2}(c_3-2)^2\,\rho^2-\frac{\xi}{6} \left[6c_1^2-12 c_3 +(c_3-2)^3\xi^2 \right]\,\rho^3+\mathcal{O}\left(\rho^4\right)\right]\,,\nonumber\\[4mm]
a_3&=&c_3+3\left(c_1^2-2c_3\right)\,\rho^2 - 4\left[6 c_3 + c_1^2 \left( -3 + \xi (c_3-2) \right)  \right]\,\rho^3+\mathcal{O}\left(\rho^4\right)\,,\nonumber \\[4mm]
h & = & h_0 - \left[ \qc +  \frac{2\,k\,\Qf\,\left(2\pi\alpha'\right)^2}{n_{\rm f}} \left( c_3^2-c_1^2(c_3-2)  \right) \right]\, \rho + \mathcal{O}(\rho^2)\,,
\end{eqnarray}
where we have  solved for the color charge using (\ref{D3charge}). In order for the solution to have the same asymptotics as in \cite{Benini} we will require $h_0=0$.

The IR at $\rho\to-\infty$ is more complicated. The equations do not admit a simple expansion in powers of $\rho$ but a double exponential-polynomial expansion is needed:
\begin{eqnarray}\label{eq.IRexpansionD3D7}
a_1&=&\tilde{c}_1\,e^{2\xi\, \rho} - \frac{\xi}{6}\tilde{c}_1\tilde{c}_3 \,e^{2\xi\, \rho} \left[3 \, e^{2\rho} (1-6\rho)^{1/3} + (3e)^{1/3}\, \Gamma \left( \frac{1}{3} , \frac{1}{3}-2\rho \right)\right]+\mathcal{O}\left(e^{6\rho}\right)\nonumber\\[4mm]
a_3&=&\tilde{c}_3\,e^{2\rho}\left(1-6\rho\right)^{1/3} + \frac{\tilde{c}_1^2}{2} \,e^{4\xi\, \rho}   \left[ 1 -2\xi e^{2(1-2\xi)\rho}\left(\frac{e^{2\xi-1}\left(1-6\rho\right)}{3(2\xi-1)^2}\right)^{1/3} \Gamma \left( \frac{2}{3} , \frac{1-6\rho}{3}(2\xi-1) \right) \right]\nonumber\\[2mm]
&+&\mathcal{O}\left(e^{6\rho}\right)\nonumber \\[4mm]
h & = & \tilde h_0 + \left( \frac{-1}{3e} \right)^{2/3} \frac{\qc}{2^{4/3}} \left[ \Gamma \left( \frac{1}{3} , - \frac{2}{3} + 4 \rho \right) -  \Gamma \left( \frac{1}{3} , - \frac{2}{3}  \right) \right]  + \cdots
\end{eqnarray}
Notice that the instanton charge, and along with it the number of color branes in the UV, only depends on the UV asymptotic behavior of these functions, in particular on the combination
\bea
\label{eq.instantonchargeD3D7}
k  &=& \frac23j\left(j+1\right)\left(2j+1\right) \,\, \lim_{\rho\to0} \frac14\left[a_1^2(2-a_3)+a_3^2\right] \nonumber \\[2mm]
&=& \frac16j\left(j+1\right)\left(2j+1\right) \left[c_1^2(2-c_3)+c_3^2\right] \,,
\eea
where we have assumed an SU($\nff$) instanton built out of a single irreducible representation of SU(2). 

We now come back  to the case $\xi=1$. The full solution must be found numerically with appropriate boundary conditions. A generic selfdual configuration will depend on two parameters, for example the values of the UV constants $c_1, c_3$. We expect one combination of them to correspond to the instanton size. We will fix the other combination by requiring that the  instanton density vanish in the UV, i.e.~at 
$\rho=0$, since we want the instanton density to be supported away from the Landau pole. This requirement selects either $c_1=0$ or $c_3=2$. In the first case the equations can be integrated exactly to 
\begin{eqnarray}\label{pseudoabelian}
a_1&=&a_2\,\,=\,\,0\,,\nonumber\\[2mm]
a_3&=&c_3\,e^{2g}\,=\,c_3\,\left(1-6\rho\right)^{1/3}\,e^{2\rho}\,.
\end{eqnarray}
This particular solution is effectively Abelian and does not give rise automatically to a quantized instanton number so we will not consider it further.\footnote{There is another exact Abelian solution given by $a_2=a_3=0$ and $a_1=e^{2\rho}$ that, albeit regular, will also not be considered further.}

Instead we will fix $c_3=2$, which automatically implies that the instanton charge is quantized in terms of $j$. This UV condition has two effects. First, together with the requirement $h_0=0$, it implies that the UV behavior of the warp factor is
\be
h = -\Qc \, \rho + \mathcal{O}(\rho^2)\,,
\ee
with 
\be
\Qc = \qc +   \frac{8\,k\,\Qf\,\left(2\pi\alpha'\right)^2}{n_{\rm f}}\,.
\label{wemay2}
\ee
This is the analog of Eq.~\eqq{wemay} and it shows that the charge-$k$ instanton contribution to the total color charge is the second term on the right hand side. The quantization condition of the color charge is 
\be\label{eq.qcvalue}
\Qc = \frac{(2\pi\ell_s)^4}{2\pi\,V_5} \lambda \,, 
\ee
with $V_5=\int \omega_5$ the volume of the SE manifold.  Together with \eqq{D7quant} and the relation between volumes $4V_5=V_3 V_1$, with $V_3$ the volume of the three-cycle wrapped by each stack of branes (in this case $V_3=2\pi^2$ for a three-sphere),  implies that we may rewrite Eq.~\eqq{wemay2} as \eqq{wemayindeed}, 
where in this case $\ncc$ is the number of color D3-branes that are not dissolved inside the flavor D7-branes and $k\,\nf/n_{\rm f} $ is the number of  D3-branes carried by the instantons. We thus conclude that at the UV one finds simply the expected D3-D7 solution with total color charge given by the sum of the dissolved plus the undissolved charges, as for the D2-D6 system. Also in parallel with that case, if $\qc \neq 0$ then the IR solution is the same as in the D3-D7 without instanton but with $\ncc$ as the effective number of color branes. This can be seen by expanding the $\Gamma$-function in \eqq{eq.IRexpansionD3D7} for $\rho \to -\infty$: 
\be
h =  \frac{ \qc }{ 3^{2/3} 2^{8/3} }  \frac{ e^{-4\rho} }{ (-\rho)^{2/3} } + \cdots \,.
\ee
This is precisely the IR behavior of the D3-D7 geometry of \cite{Benini} with color  charge $\qc$. Thus the solution is singular in the IR, as in that reference. If 
$\qc=0$ the IR behavior of the warp factor is milder,
\be
h = - 6 \tilde c_3^2 \, \Qf \, \rho + \cdots \,,
\label{milder}
\ee
but the solution is still singular, as illustrated by the fact that scalar curvature diverges as $\rho \to -\infty$.

\begin{figure}[t]
\begin{center}
\begin{subfigure}{.45\textwidth}
\includegraphics[width=\textwidth]{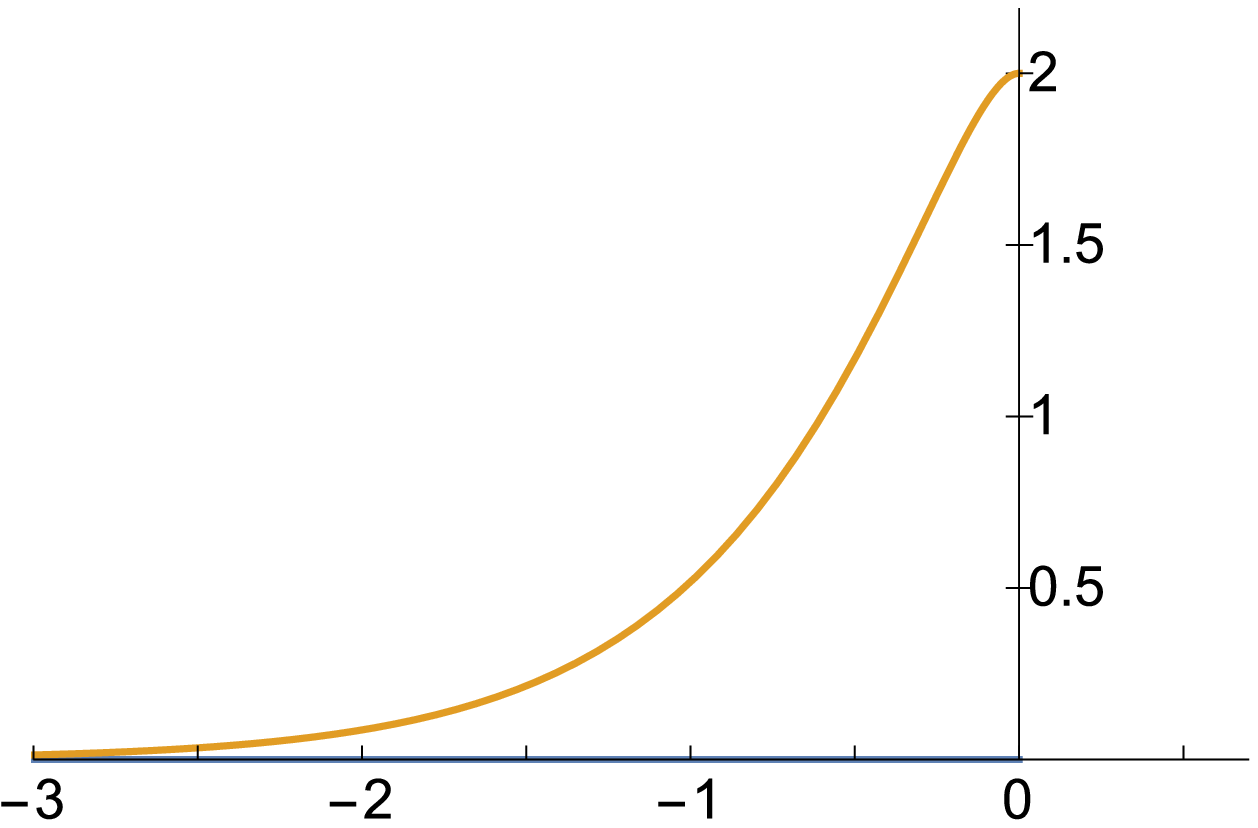} 
\put(-48,135){$a_{1,3}$}
\put(-5,5){$\rho$}
\caption{$c_1=0$}
\end{subfigure}\hfill
\begin{subfigure}{.45\textwidth}
\includegraphics[width=\textwidth]{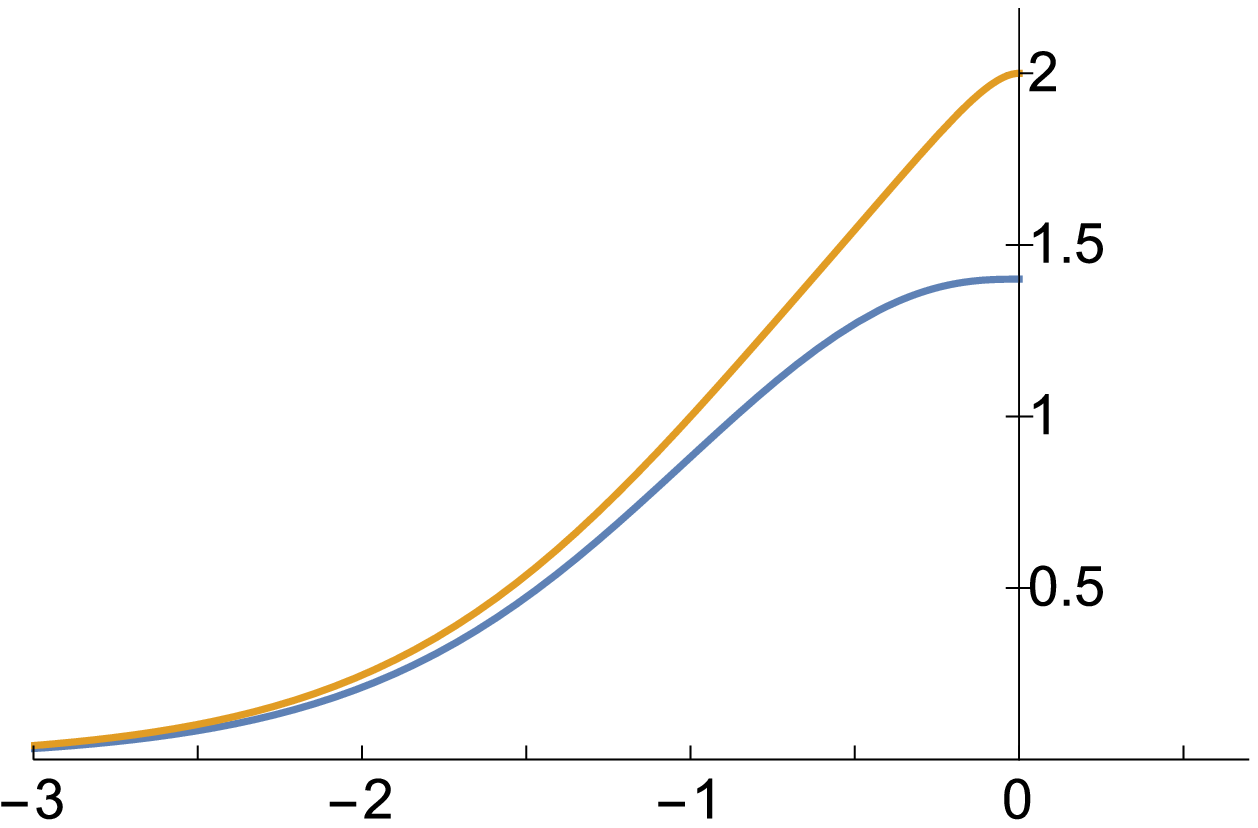} 
\put(-48,135){$a_{1,3}$}
\put(-5,5){$\rho$}
\caption{$c_1=1$}
\end{subfigure}
\begin{subfigure}{.45\textwidth}
\includegraphics[width=\textwidth]{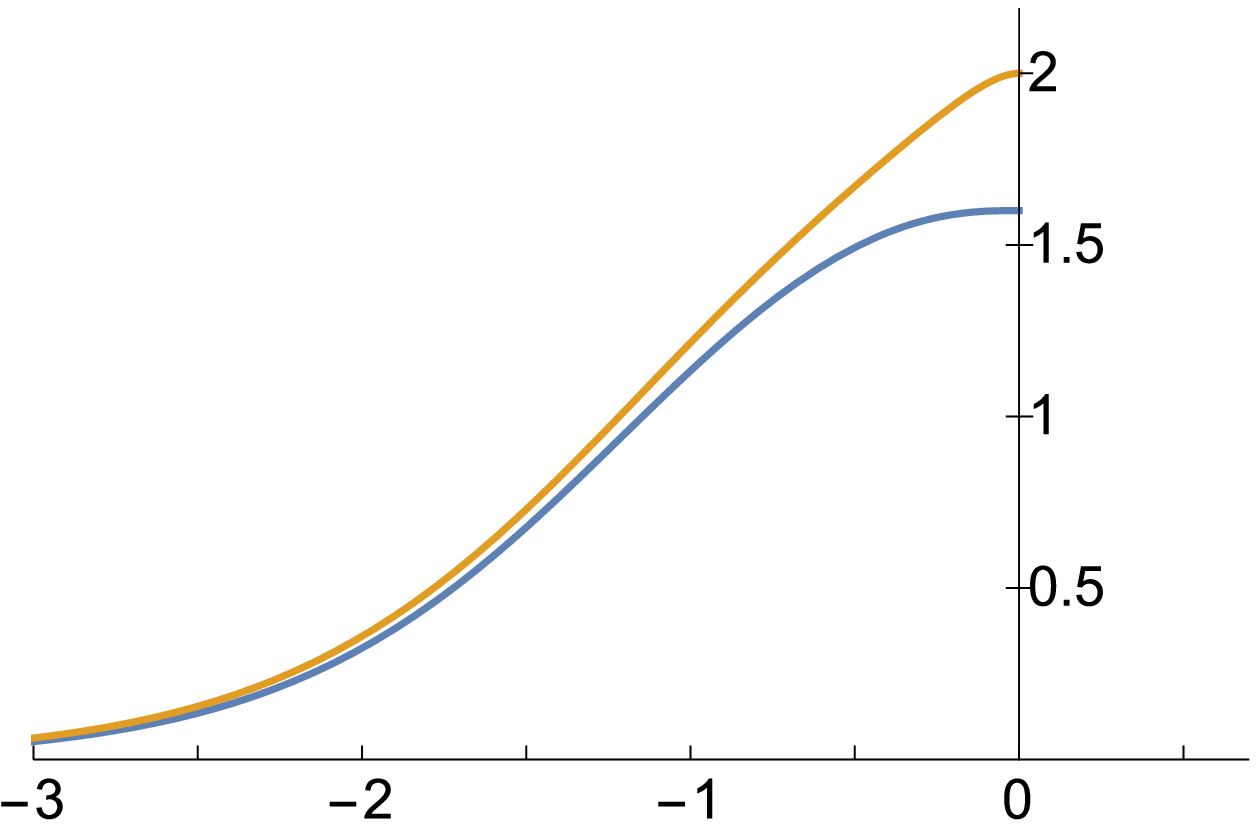} 
\put(-48,135){$a_{1,3}$}
\put(-5,5){$\rho$}
\caption{$c_1=1.4$}
\end{subfigure}\hfill
\begin{subfigure}{.45\textwidth}
\includegraphics[width=\textwidth]{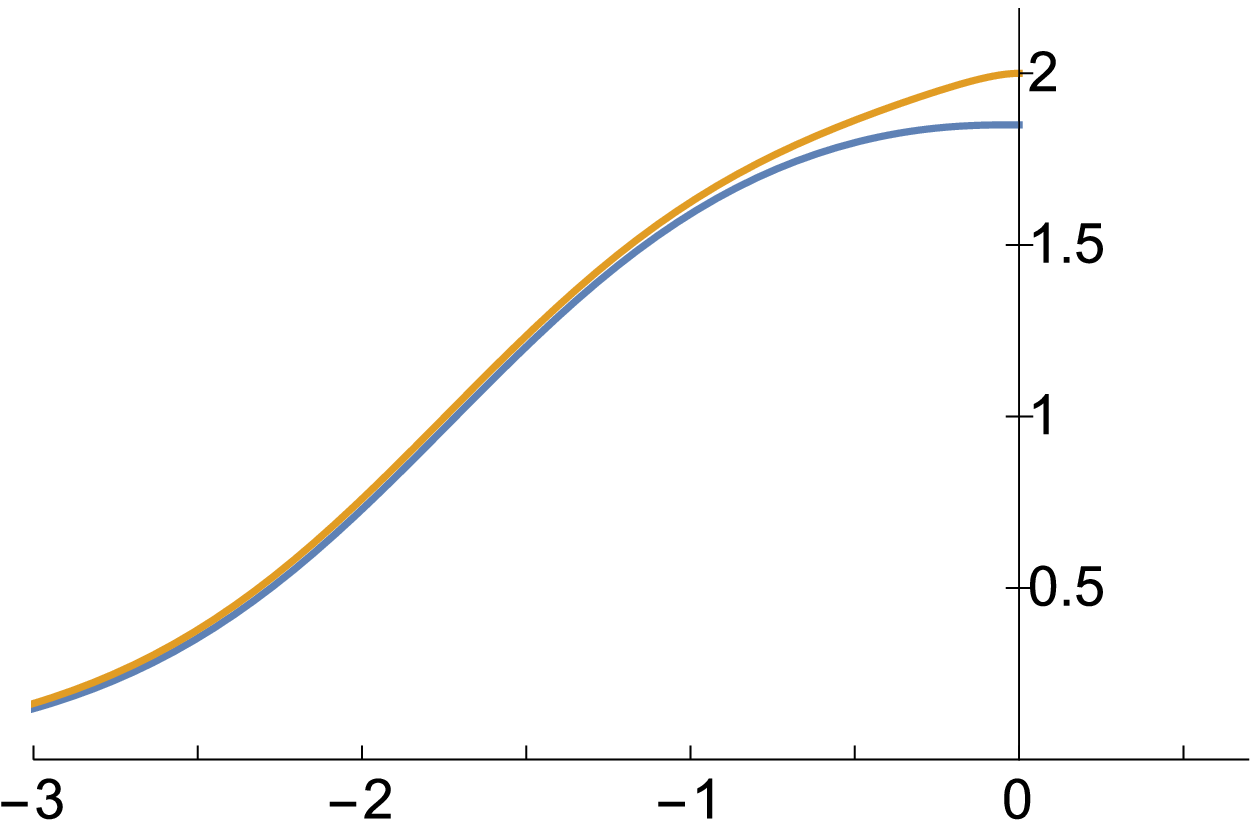} 
\put(-48,135){$a_{1,3}$}
\put(-5,5){$\rho$}
\caption{$c_1=1.9$}
\end{subfigure}
\caption{Gauge field potentials $a_1=a_2$ (blue) and $a_3$ (orange) for $c_3=2$ and different values of $c_1$ for the instanton solution in the case of SYM theory.}\label{fig.vecpotsnp1}
\end{center}
\end{figure}

\begin{figure}[t!]\begin{center}
\begin{subfigure}{.45\textwidth}
\includegraphics[width=\textwidth]{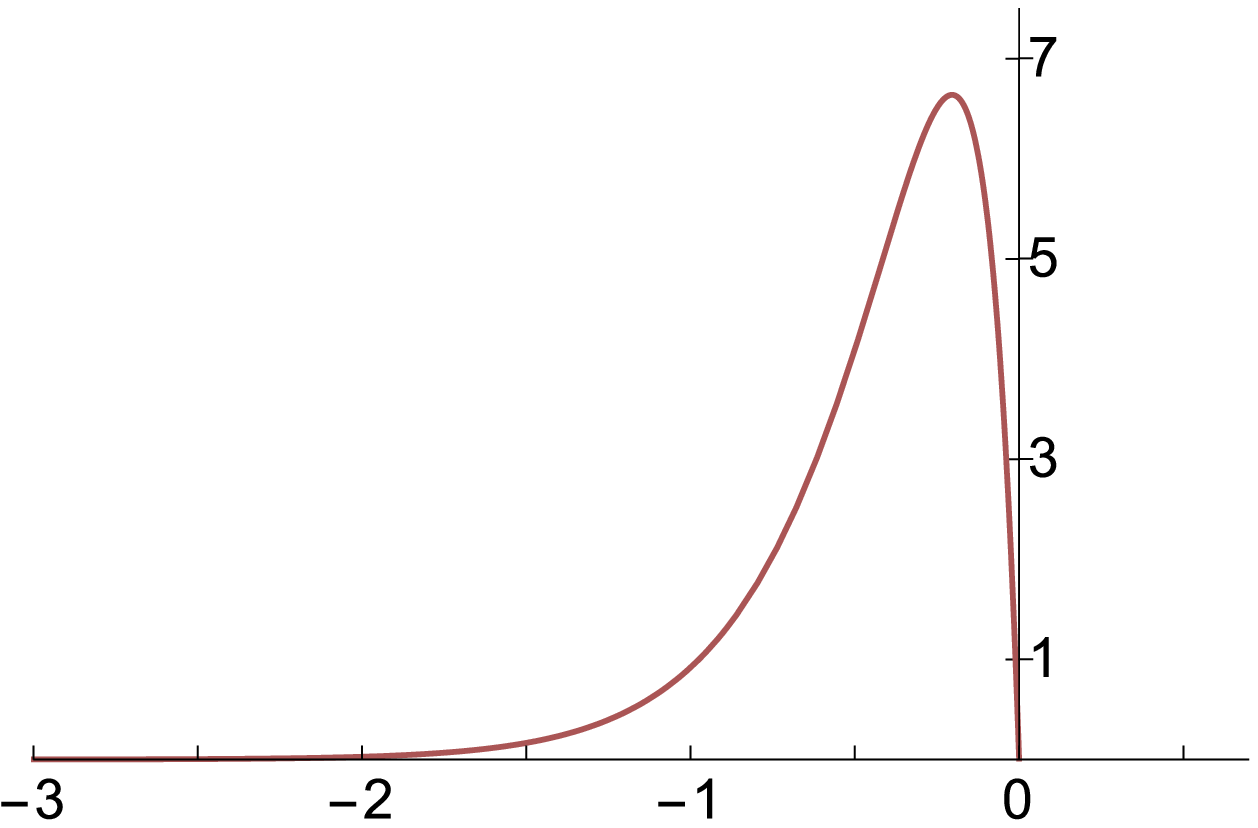} 
\put(-46,130){$\Psi$}
\put(-5,5){$\rho$}
\caption{$c_1=0$}
\end{subfigure}\hfill
\begin{subfigure}{.45\textwidth}
\includegraphics[width=\textwidth]{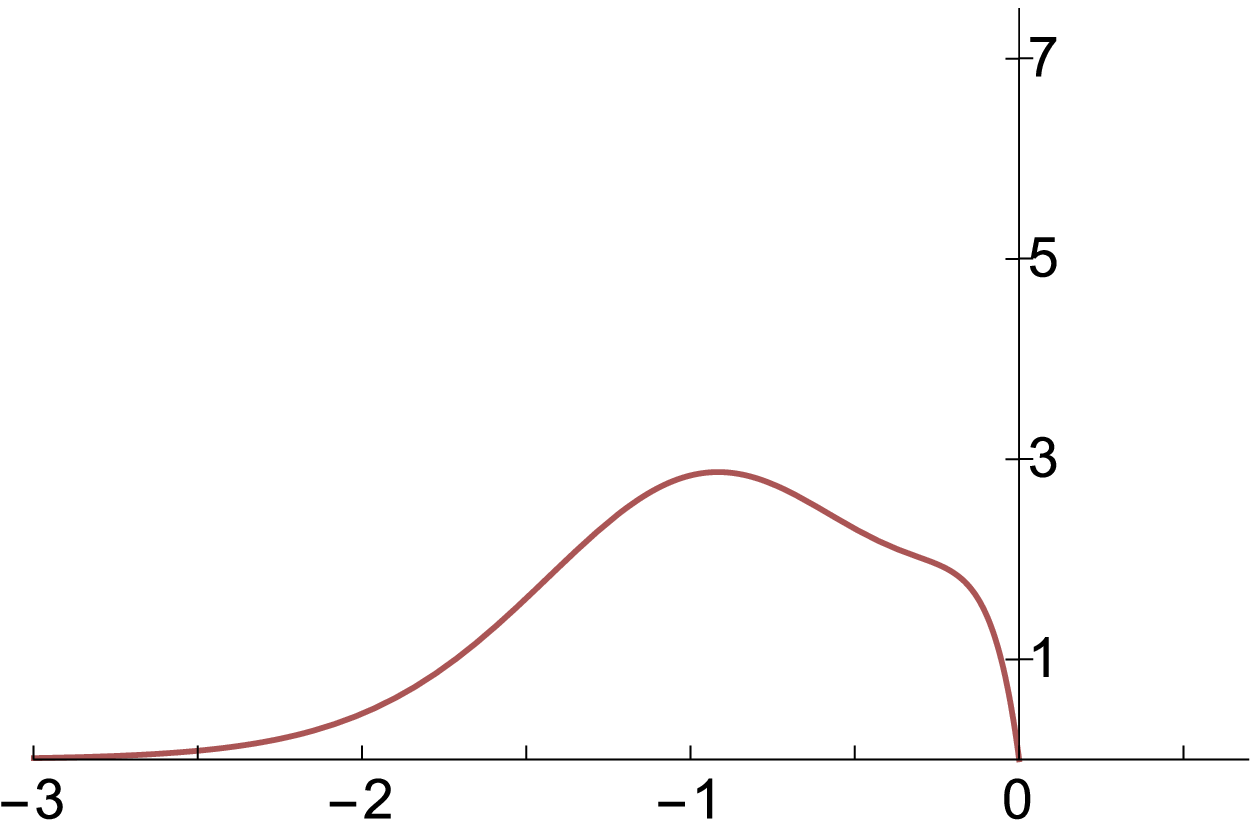} 
\put(-46,130){$\Psi$}
\put(-5,5){$\rho$}
\caption{$c_1=1$}
\end{subfigure}
\begin{subfigure}{.45\textwidth}
\includegraphics[width=\textwidth]{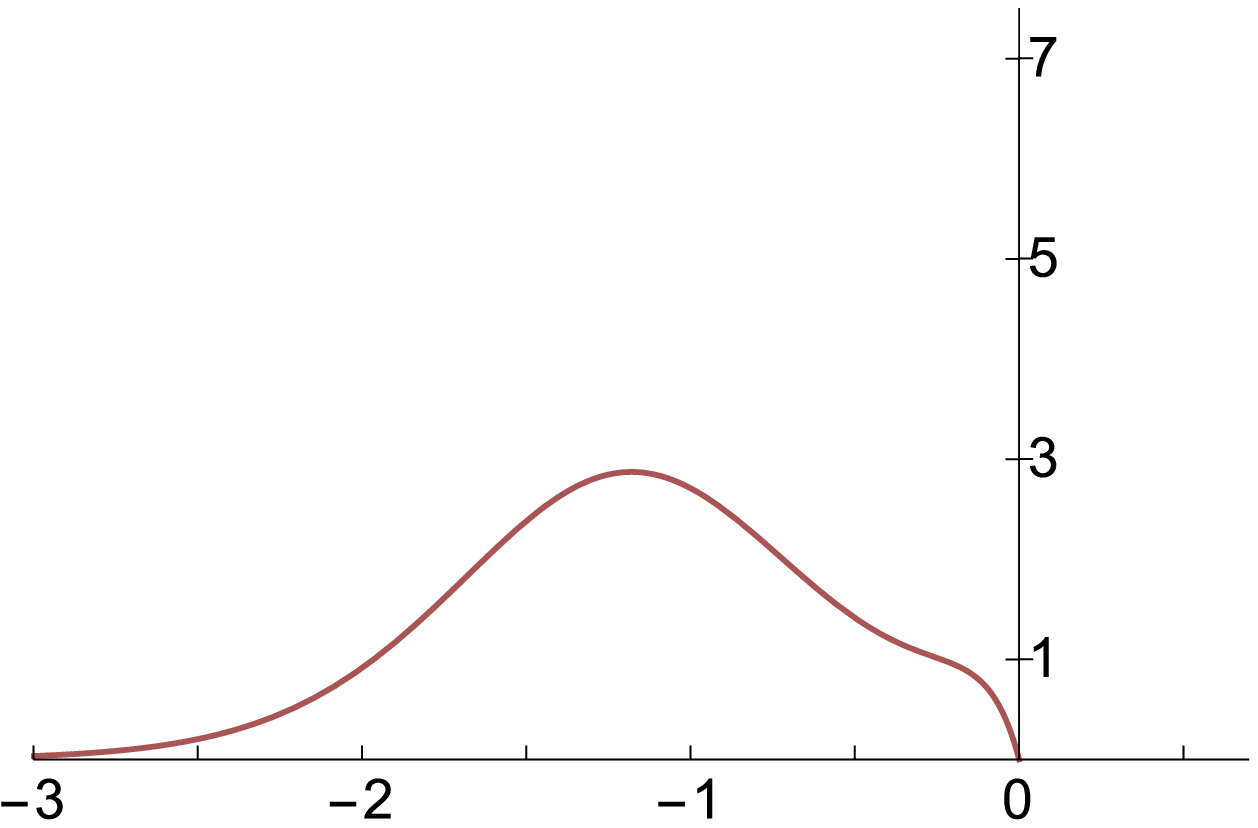} 
\put(-46,130){$\Psi$}
\put(-5,5){$\rho$}
\caption{$c_1=1.4$}
\end{subfigure}\hfill
\begin{subfigure}{.45\textwidth}
\includegraphics[width=\textwidth]{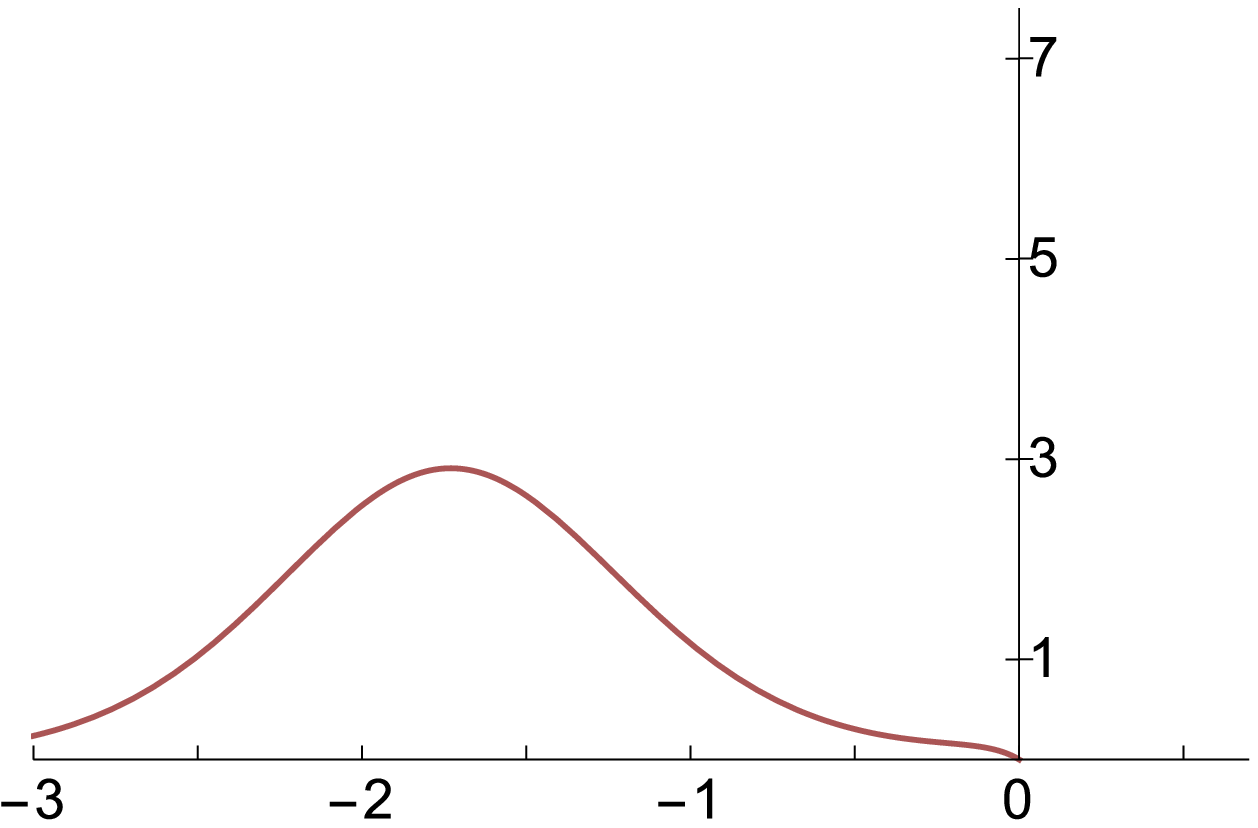} 
\put(-46,130){$\Psi$}
\put(-5,5){$\rho$}
\caption{$c_1=1.9$}
\end{subfigure}
\caption{Instanton density $\Psi (\rho)$ for $c_3=2$ and different values of $c_1$ for the instanton solution in the case of SYM theory.}\label{fig.psinp1}
\end{center}\end{figure}

\begin{figure}[t!]\begin{center}
\begin{subfigure}{.42\textwidth}
\includegraphics[width=\textwidth]{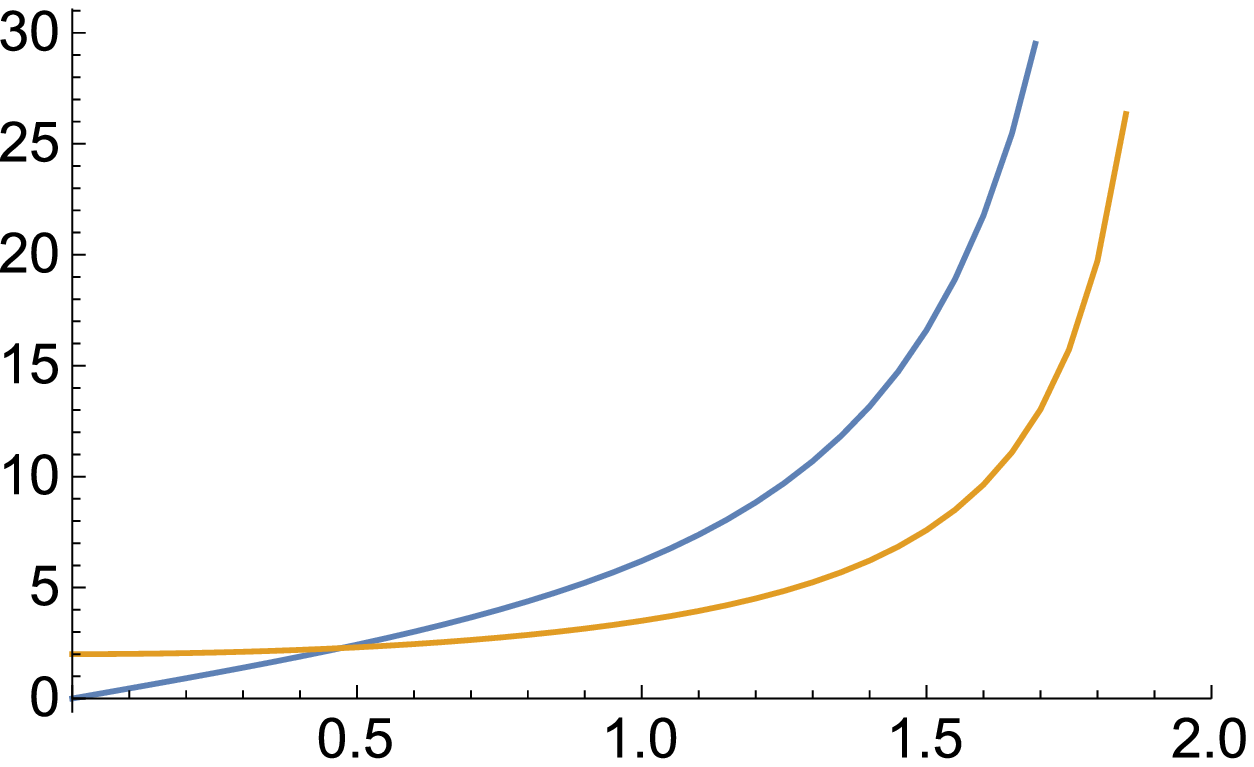} 
\put(-190,125){$\tilde c_{1,3}$}
\put(5,4){$c_1$}
\end{subfigure}
\qquad\qquad
\begin{subfigure}{.42\textwidth}
\includegraphics[width=\textwidth]{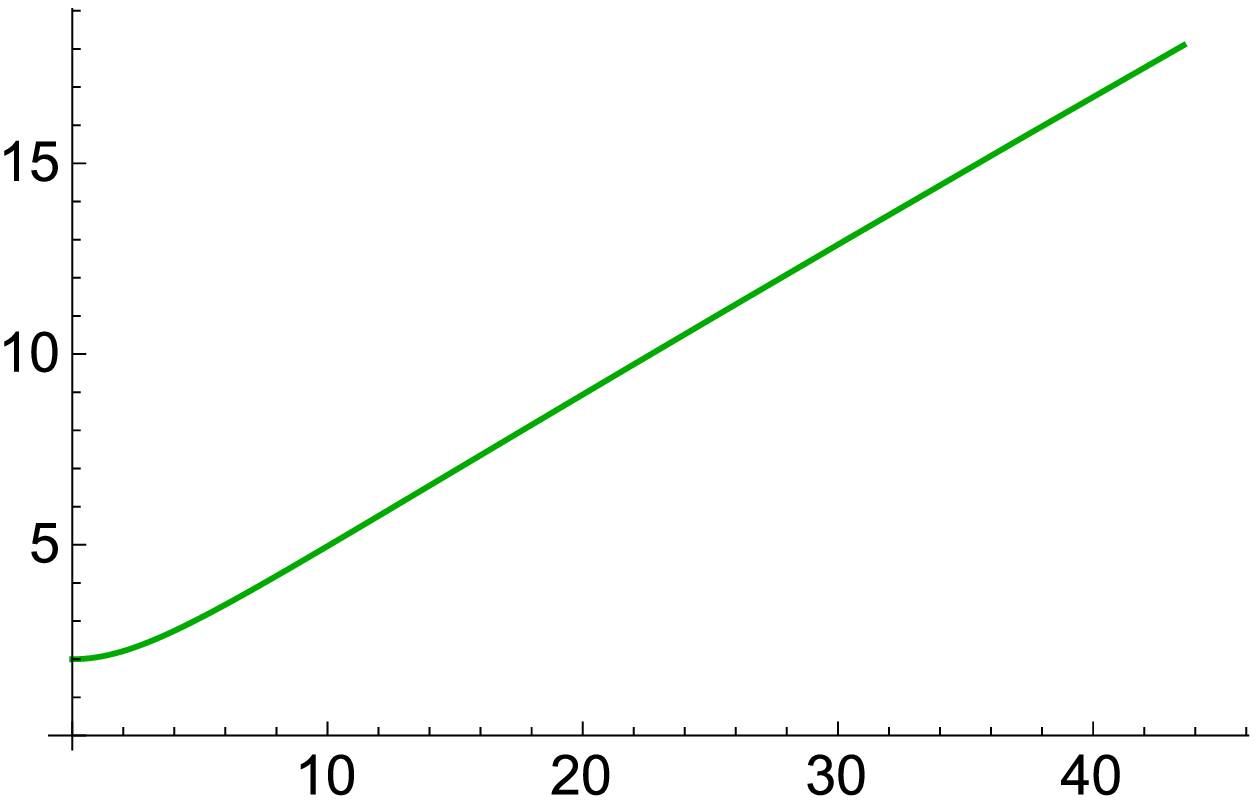} 
\put(-185,125){$\tilde c_{3}$}
\put(5,4){$\tilde c_1$}
\end{subfigure}
\caption{(Left) The IR parameters $\tilde c_1$ (blue) and $\tilde c_3$ (orange) as a function of the UV parameter $c_1$ for the instanton solution in the case of SYM theory. (Right) Relationship between the IR parameters.}\label{fig.c1c3np1}
\end{center}\end{figure}

The second effect of the UV condition $c_3=2$ is that it  forces a relation between the integration constants in the IR and it leaves us with a one-parameter family of solutions labelled by $c_1$. As we will now see, this parameter plays the role of the size of the instanton,  $\Lambda$. There is a symmetry $c_1 \to - c_1$ inherited from the $a_1 \to -a_1$ symmetry in the equations of motion, so we will consider only $c_1 \geq 0$. When $c_1=0$ we have that $a_1$ vanishes exactly and we are back to the Abelian solution (\ref{pseudoabelian}). The values of the IR parameters $\tilde c_1$, $\tilde c_3$ and $\tilde h_0$ are determined as functions of $c_1$. Since $\tilde c_1$ and  $\tilde c_3$ diverge as $c_1 \to 2^-$ we will only consider $0< c_1 < 2$. We provide the numerically-generated profiles for $a_1$ and $a_3$ in Fig.~\ref{fig.vecpotsnp1} for selected values of the constant of integration $c_1$. 
Similarly, in Fig.~\ref{fig.psinp1} we provide the radial profiles  for the  instanton density $\Psi(\rho)$ as determined from Eq.~\eqref{Psi}. Inspection of these  plots shows that for low values of $c_1$ the instanton charge is  located closer to the UV than for the cases with $c_1\sim 2$. At intermediate values some structure seems to be present, but this might be an artifact of the radial coordinate used to integrate the equations of motion.
The IR  parameters $\tilde c_1$ and $\tilde c_3$ as a function of $c_1$ are shown in Fig.~\ref{fig.c1c3np1}(left). They start with the values  $\tilde c_1=0$ and $\tilde c_3=2$ and then grow monotonically with $c_1$ until $c_1=2$ is reached. At this point  the  IR parameters  diverge. In Fig.~\ref{fig.c1c3np1}(right) we show the parametric dependence of $\tilde c_3$ as a function of $\tilde c_1\geq0$ (this range is inherited from the $a_1\to -a_1$ symmetry we have used above to set $c_1>0$). We observe that at large values of the IR parameters the relationship between the two is approximately linear. 

As in the D2-D6 case, we could consider an SU($\nff$) instanton built out of several irreducible representations of SU(2) of different dimensions. In this case the total instanton charge and the total instanton densities would take the form \eqq{total} (with $r$ replaced by $\rho$), where each $\Psi_n(\rho)$ would be determined by a corresponding value of the UV integration constant, $c_1^n$. This general solution would give rise to an RG flow that would gradually transition at the different scales associated to the $c_1^n$ from the D3-D7 solution with charges $\Qc, \Qf$ in the far UV to either the D3-D7 solution with charges $\qc, \Qf$ or the solution \eqq{milder} in the deep IR.  

\subsection{Klebanov--Witten quiver theory}\label{sec.otherquivers4d}

The so called ${\rm T}^{1,1}$ is a SE manifold with a well known dual gauge theory: the Klebanov--Witten quiver. Its metric, normalized to $R_5=20$, which can be written as
\begin{equation}
\dd s_5^2\,=\,\frac16\sum_{i=1}^2\left(\dd \theta_i^2+\sin^2\theta_i\dd\varphi_i^2\right)+\frac19\left(\dd \psi+\sum_{i=1}^2\cos\theta_i\dd\varphi_i\right)^2 
\end{equation}
with the ranges 
\be
0\le\theta_i<\pi \sac 0\le\varphi_i<2\pi \sac 0\le\psi<4\pi \,,
\ee
describes a U(1) bundle over ${\rm S}^2\times{\rm S}^2$. There are two types of BPS embeddings for D7-branes in this geometry.

\subsubsection{Chiral embedding}

We first consider chiral embeddings in which $\theta_i$ and $\varphi_i$ are constant for either $i=1$ or $i=2$. Due to the $\mathbb{Z}_2$ symmetry interchanging the spheres the two  choices are equivalent.  In terms of left-invariant forms normalized as in  Eq.~\eqq{normalized} the induced metric on the D7-branes is
\begin{equation}
\dd s^2\,=\,h^{-1/2}\dd x_{1,3}^2+h^{1/2}\,e^{2f}\left(\dd\rho^2+e^{2g-2f}\,\frac23\,\left(\omega_1^2+\omega_2^2\right)+\frac49\,\omega_3^2\right)\,.
\end{equation}
Notice that we have again a radial-dependent squashing of the fiber with respect to the base. The solution for the metric components is as in (\ref{D3D7smeared}). (Anti)selfduality requires the different gauge-potential components to verify
\begin{eqnarray}
a_1'&=&\frac{3\sigma}{2}\left(2a_1-a_2\,a_3\right)\,,\nonumber\\[2mm]
a_2'&=&\frac{3\sigma}{2}\left(2a_2-a_3\,a_1\right)\,,\\[2mm]
a_3'&=&\sigma\,e^{2f-2g}\,\left(2a_3-a_1\,a_2\right)\,.\nonumber
\end{eqnarray}
The numerical coefficient on the right-hand-side of the equations is different compared to that  for $\rm{S}^5$. This translates into a difference in the powers of the radial expansions and in the dimension of the dual operator and the asymptotic expansions \eqref{eq.UVexpansionD3D7} and \eqref{eq.IRexpansionD3D7}, which must be read with $\xi=3/2$. Repeating the numerical analysis performed for the $\rm{S}^5$ case we find no qualitative differences for the quantities plotted in  
Figs.~\ref{fig.psinp1} and \ref{fig.c1c3np1}. 

\subsubsection{Non-chiral embedding}

The other type of supersymmetric embeddings are dubbed non-chiral and correspond to
\begin{equation}
\theta_1\,=\,\theta_2\,=\,\theta\,,\qquad\qquad\qquad\qquad\varphi_1\,=\,\varphi_2\,=\,\varphi\,.
\end{equation}
The cycle wrapped by the branes is thus 
\begin{eqnarray}
\dd s_3^2&=&\frac13\left(\dd \theta^2+\sin^2\theta\dd\varphi^2\right)+\frac19\left(\dd \psi+2\cos\theta\dd\varphi\right)^2\nonumber\\[2mm]
&=&\frac13\left(\dd \theta^2+\sin^2\theta\dd\varphi^2\right)+\frac49\left(\dd \hat{\psi}+\cos\theta\dd\varphi\right)^2\nonumber\\[2mm]
&=&\frac43\,\left(\omega_1^2+\omega_2^2\right)+\frac{16}{9}\,\omega_3^2
\end{eqnarray}
where we have defined $2\hat{\psi}=\psi$ and as a result we are dealing with a (squashed) $\mathbb{RP}^3$ instead of a sphere, which must be taken into consideration when computing the instanton number. To construct a self-dual solution we need to solve
\begin{eqnarray}
a_1'&=&\frac{3\sigma}{4}\left(2a_1-a_2\,a_3\right)\,,\nonumber\\[2mm]
a_2'&=&\frac{3\sigma}{4}\left(2a_2-a_3\,a_1\right)\,,\\[2mm]
a_3'&=&\sigma\,e^{2f-2g}\,\left(2a_3-a_1\,a_2\right)\,.\nonumber
\end{eqnarray}
The numerical coefficient on the right-hand-side of the equations is different compared with those for $\rm{S}^5$ or the chiral embedding described above. This translates into a difference in the powers of the radial expansions and in the dimension of the dual operator and the asymptotic expansions \eqref{eq.UVexpansionD3D7} and \eqref{eq.IRexpansionD3D7}, which must be read with $\xi=3/4$. Repeating the numerical analysis performed for the $\rm{S}^5$ case we find no qualitative differences for the quantities plotted in  Figs.~\ref{fig.psinp1} and \ref{fig.c1c3np1}.  

\section{Discussion}\label{sec.discussion}

We have constructed the supergravity duals of states on the Higgs branches of three- and four-dimensional  
$\mathcal{N}=1$ supersymmetric gauge theories with flavor. On the gravity side the system consists of $\nc$ color D2-branes and $\nf$ flavor D6-branes in the three-dimensional case, and $\nc$ color D3-branes and $\nf$ flavor D7-branes in the four-dimensional case.

The most dramatic effect of the Higgsing on the gauge theory side is that the effective number of colors decreases with the energy scale from $\nc$ in the UV to $\ncc < \nc$ in the IR. On the gravity side the Higgsing is encoded in the dissolution of $\nc-\ncc$ of the color branes inside the flavor branes. From the viewpoint of the flavor branes, the dissolved color branes appear as an instanton configuration of the non-Abelian gauge fields living on their worldvolume. In order to see the effect of the Higgsing on the gravity side we have included the backreaction of the instanton on the supergravity fields. The result is a cascading-like solution in which the effective number of color branes, measured by the  flux of the appropriate Ramond-Ramond form, decreases along the holographic direction. 

Our solutions are supersymmetric as expected from the fact that  states on the Higgs branch preserve the supersymmetries of the gauge theory. In fact, the VEVs of the operators that parametrize the Higgs branch are exact moduli in the gauge theory. Moreover, different operators can acquire different VEVs, thus breaking different parts of the gauge group at  different scales. On the gravity side these features are  reflected in the fact that instantons of different sizes can be essentially superposed, and in the fact that these sizes  are  arbitrary parameters in our solutions. 

The picture that emerges on the gravity side is conceptually very simple. Consider first the case of massless quarks. If the Higgsing is only partial, i.e.~if $\ncc \neq 0$, then the solutions interpolate from the usual backreacted solutions with $\nc$ color branes and $\nf$ flavor branes in the UV, to the same solution in the IR but with $\nc$ replaced by $\ncc$. In the D2-D6 case the UV solution is the usual D2-brane solution, since the D6-branes give subleading contributions in this limit, whereas the IR solution is  AdS$_4$ \cite{Faedo}. In the D3-D7 case the UV is a singular geometry due to the presence of a Landau pole in the gauge theory, and the IR geometry is a 
 ``logarithmically-corrected'' (and singular) AdS$_5$ solution \cite{Benini}. If instead the Higgsing is complete and $\ncc = 0$, then the UV geometries remain  the same  but the IR geometries are completely modified and they turn out to be singular both in the D2-D6 and in the D3-D7 cases. The flow from the far UV to the deep IR geometries can either take place through  a single transition if a single VEV on the Higgs branch is turned on, or through several transitions if several VEVs are turned on. On the gravity side a transition takes place each time that the holographic radial coordinate reaches the characteristic size of one of the instantons in the solution. In the D2-D6 case this has the interesting consequence that, for solutions with multiple instantons of appropriate sizes, there exist several distinct regions in which the geometry is approximately AdS$_4$ separated from one another by the transitions above. On the gauge theory side this corresponds to several energy ranges in which the  theory exhibits quasi-conformal or ``walking'' dynamics. 
 
 The main modification of the picture above introduced by a non-zero quark mass is that the flavor branes terminate at a radial position proportional to the quark mass. As a consequence,  if the breaking of the gauge group is only partial then the solution below the scale set by the quark mass  is the usual solution corresponding to $\ncc$ color branes. If instead the breaking is complete, then $\ncc=0$ and the solution below the scale set by the quark mass is simply flat space. 
 
\section*{Acknowledgements}

We are grateful to Prem Kumar, Alfonso Ramallo and Diego Rodriguez-Gomez for discussions. AFF, DM and CP are supported by grants MEC FPA2013-46570-C2-1-P, MEC FPA2013-46570-C2-2-P, MDM-2014-0369 of ICCUB, 2014-SGR-104, 2014-SGR-1474, CPAN CSD2007-00042 Consolider-Ingenio 2010, and ERC Starting Grant HoloLHC-306605. JT is supported by the Advanced ARC project ``Holography, Gauge Theories and Quantum Gravity'' and by the Belgian Fonds National de la Recherche Scientifique FNRS (convention IISN 4.4503.15).

\appendix

\section{An instanton shell}\label{appendixA}

In the bulk of the paper we have considered instanton solutions centered at the origin of the four-dimensional space transverse to the color branes. One may also consider solutions in which the instanton center is at an arbitrary point in this space. In the gauge theory this corresponds to turning on a vacuum expectation value for some of the adjoint scalars. These type of  solutions would still be supersymmetric, but generically they would  break some rotational symmetries and therefore they are not captured by our ansatz. However, since the instantons are mutually BPS, one may consider a distribution of a large number of them arranged in such a way that rotational symmetry is preserved if the distribution is treated as continuous. In this section we will work in this approximation, which of course is valid only at length scales that are large compared to the inter-instanton distance. For simplicity, we will assume that the four-dimensional space is $\mathbb{R}^4$, as in e.g.~the case of massless quarks in three-dimensional SYM. 

 It can be seen that a multi-instanton solution for the gauge group SU(2) can be simply obtained using an ansatz due to 't Hooft (see also \cite{Jackiw}) in which the gauge potential reads
\begin{equation}
a_m\,=\,\frac12\,\sigma_{mn}\,\partial_n\log\phi\,,
\end{equation}
with $\sigma_{mn}$ the $2\times2$ matrix representation of the Lorentz generators in the four-dimensional Euclidean space, whose properties can be found elsewhere. A selfdual configuration can be obtained by solving the Laplace equation for the potential 
\begin{equation}
\frac{\Box\phi}{\phi}\,=\,0\,.
\end{equation} 
The single-instanton case corresponds to the solution (in spherical coordinates)
\begin{equation}
\phi\,=\,1+\frac{\Lambda^2}{r^2}\,,
\end{equation}
while for $n$ instantons centered at the positions $\vec z_i$ the solution is 
\begin{equation}\label{multipotential}
\phi\,=\,1+\sum_{i=1}^n\frac{\Lambda_i^2}{\left(\vec z- \vec z_i\right)^2}\,,
\end{equation}
where $\vec z$ are Cartesian coordinates in $\mathbb{R}^4$. 
Notice that when all the instantons are centered at the same position the solution collapses to that of a single instanton of size $\Lambda^2=\sum_i\Lambda_i^2$. From this potential the instanton density is computed as
\begin{equation}
{\rm tr}\left(F_{mn}F^{mn}\right)\,=\,-\frac12\,\Box\Box\log{\phi}\,,
\end{equation}
and its integral, counting the instanton number, gets a contribution from each pole in (\ref{multipotential}). We can consider a large number of instantons all located at a distance $r_0$ from the origin and take the continuous limit, restoring in this way the spherical symmetry. The problem is thus analogous to computing the electrostatic potential for a charged sphere. The potential outside the ball is as if all the charge were at the origin, while inside the ball it is constant:
\begin{eqnarray}
\phi&=&\left\{\begin{array}{ll}
		1+\frac{\Lambda^2}{r_0^2}  & \qquad{\rm if } \,\,\, r\le r_0 \\[4mm]
		1+\frac{\Lambda^2}{r^2}&\qquad{\rm if } \,\,\, r> r_0\,.
	\end{array}
\right.
\end{eqnarray}
The resulting instanton density is however not continuous and indeed vanishing for $r<r_0$. As a consequence, the color charge density \eqq{QCC}, which is the integral of the instanton density,  is continuous but not differentiable. This is expected since these features occur at $r=r_0$ and we know that the solution is only trustable  at large distances from the shell. Moreover, the total instanton charge is 
\begin{equation}
k\,=\,-\frac{1}{16\pi^2}\int*_4\Box\Box\log\phi\,=\,\frac{\Lambda^4\left(3r_0^2+\Lambda^2\right)}{\left(r_0^2+\Lambda^2\right)^3}\,.
\end{equation}
The result on the right-hand side is not quantized and, in fact, it is always smaller than 1 if $r_0 >0$. These features are not surprising given that all the multi-instanton nature of the solution \eqq{multipotential} is encoded in the regions near each of the poles, which are discarded here due to the smearing and the continuous approximation. 


The form of the supergravity solution obtained from the backreaction of this instanton shell is extremely simple. Outside the shell, at $r>r_0$, the solution is exactly the same as those in the bulk of the paper for an instanton of size $\Lambda$. As $r \to r_0^+$  the supergravity fields approach finite values. The solution inside the shell is simply that of the D2-D6 system with no instanton \cite{Faedo} with charges $\qc$ and $\Qf$ with the boundary condition that the value of the fields as $r\to r_0^-$ agrees with their value as $r \to r_0^+$.


\end{document}